\documentclass[11pt]{article}
\renewcommand{\a}{\alpha}
\renewcommand{\b}{\beta}
\newcommand{\g}{\gamma}           
\renewcommand{\d}{\delta}         
\newcommand{\e}{\varepsilon}

\newcommand{\la}{\lambda}

\newcommand{\om}{\omega}         \newcommand{\OM}{\Omega}
\newcommand{\p}{\psi}             

\newcommand{\s}{\sigma}           \renewcommand{\S}{\Sigma}
         \newcommand{\T}{\Theta}
           \newcommand{\F}{{\Phi}}
\newcommand{\vf}{{\varphi}}
\newcommand{\x}{\xi}              

\newcommand{\z}{\zeta}

\newcommand{\eps}{{\epsilon}}

\newcommand{\ca}{{\cal A}}
\newcommand{\cb}{{\cal B}}
\newcommand{\cc}{{\cal C}}
\newcommand{\cd}{{\cal D}}

\newcommand{\cf}{{\cal F}}

\newcommand{\cl}{{\cal L}}

\newcommand{\cs}{{\cal S}}

\newcommand{\ct}{{\cal T}}

\newcommand{\cx}{{\cal X}}

\newcommand{\be}{\begin{equation}}
\newcommand{\ee}{\end{equation}}
\newcommand{\eqn}[1]{\label{#1}\end{equation}}
\newcommand{\equ}[1]{(\ref{#1})}

\newcommand{\bea}{\begin{eqnarray}}
\newcommand{\eea}{\end{eqnarray}}
\newcommand{\eqan}[1]{\label{#1}\end{eqnarray}}

\newcommand{\ba}{\begin{array}}
\newcommand{\ea}{\end{array}}

\newcommand{\nn}{\nonumber}

\newcommand{\loco}{{\mathop{ \, \rule[-.06in]{.2mm}{3.8mm}\,}}}
\newcommand{\doubar}{{{\loco}\!{\loco}}}
\newcommand{\az}{{\bf{z}}}
\newcommand{\au}{{\bf{u}}}

\newcommand{\da}{{\dot{\alpha}}}
\newcommand{\db}{{\dot{\beta}}}
\newcommand{\dg}{{\dot{\gamma}}}
\newcommand{\dd}{{\dot{\delta}}}
\newcommand{\df}{{\dot{\varphi}}}

\newcommand{\ta}{{\mbox{\tiny{A}}}}
\newcommand{\tb}{{\mbox{\tiny{B}}}}
\newcommand{\tc}{{\mbox{\tiny{C}}}}
\newcommand{\td}{{\mbox{\tiny{D}}}}
\newcommand{\te}{{\mbox{\tiny{E}}}}
\newcommand{\tf}{{\mbox{\tiny{F}}}}
\newcommand{\tg}{{\mbox{\tiny{G}}}}
\renewcommand{\th}{{\mbox{\tiny{H}}}}
\newcommand{\ti}{{\mbox{\tiny{I}}}}
\newcommand{\tj}{{\mbox{\tiny{J}}}}
\newcommand{\tk}{{\mbox{\tiny{K}}}}
\newcommand{\tl}{{\mbox{\tiny{L}}}}
\newcommand{\tm}{{\mbox{\tiny{M}}}}
\newcommand{\tn}{{\mbox{\tiny{N}}}}

\newcommand{\vd}{\textrm{d}}
\newcommand{\vi}{\textrm{i}}

\newcommand{\fp}{\cf^\textrm{\tiny{+}}}
\newcommand{\fm}{\cf^-}%\textrm{\tiny{--}}}

\newcommand{\hfp}{\hat{F}^\textrm{\tiny{+}}}
\newcommand{\hfm}{\hat{F}^-}
\newcommand{\hOM}{\hat{\OM}}

\newcommand{\su}{{\textrm{\tiny{SU(8)}}}}
\newcommand{\dwz}{\d^{\textrm{\tiny{WZ}}}}

\newcommand{\bs}{\bar{\s}}
\newcommand{\bp}{\bar{\p}}

\newcommand{\bx}{\bar{\x}}

\newcommand{\brg}[1]{\left(#1\right)_T}

\renewcommand{\T}{\la}
\newcommand{\Sc}{{\cal{S}}}

\begin{document}

\begin{titlepage}
\begin{center}
\renewcommand{\thefootnote}{\fnsymbol{footnote}}
{%\twelve
Centre de Physique Th\'eorique\footnote{Unit\'e Propre de
Recherche 7061 }, CNRS Luminy, Case 907}

{%\twelve
F-13288 Marseille -- Cedex 9}

\vspace{3 cm}

{\huge {$N=8$ Supergravity in Central Charge Superspace}}

\vspace{2 cm}

\setcounter{footnote}{0}
\renewcommand{\thefootnote}{\arabic{footnote}}

{\bf Annam\'aria KISS\footnote{\it kiss@cpt.univ-mrs.fr} and Emmanuel LOYER\footnote{\it
loyer@cpt.univ-mrs.fr}}

\vspace{3 cm}

{\bf Abstract}\\

\end{center}
We present a geometrical description of $N=8$ supergravity, using central
charge superspace. The essential properties of the multiplet, like self-duality
properties of the vectors or the non-linear sigma model structure of the
scalars, are deduced from constraints at 0 and 1/2 canonical dimension. We also present in detail how to derive from this geometrical formulation the supergravity transformations as well as the whole equations of motion for the component fields in order to compare them with the results already known and obtained in formulations on the component level.

\vfill


\bigskip

\bigskip

\noindent \today

\noindent CPT-2001/P.4139

\bigskip

\noindent anonymous ftp : ftp.cpt.univ-mrs.fr

\noindent web : www.cpt.univ-mrs.fr

\renewcommand{\thefootnote}{\fnsymbol{footnote}}

\end{titlepage}

\setcounter{footnote}{0}
\renewcommand{\thefootnote}{\arabic{footnote}}

\tableofcontents

\newpage

%%%%%%%%%%%%%%%%%%%%%%%%%%%%%%%%%%%%%%%%%%%%%%%%%%%
\section{Introduction}
%%%%%%%%%%%%%%%%%%%%%%%%%%%%%%%%%%%%%%%%%%%%%%%%%%%

$N=8$ extended supergravity is called maximal in the sense that it does not require 
helicities larger than two (gravitons). One of its intriguing properties 
is that it admits $SU(8)$ as intrinsic gauge invariance group.
The spectrum of the multiplet looks rather
complicated at the first sight: it contains a graviton, 8 Rarita-Schwinger fields, 28 graviphotons, 56
helicity 1/2 fields and 70 helicity 0 states, which are considered to be all scalars. However, these
fields are organized in a quite simple structure which is recovered and described in a concise manner
using the geometric description of the theory in central charge superspace. One of the main objectives of
the present paper is to emphasize this feature.

The analysis of these theories began with the development of the basic structure of $SO(8)$ supergravity,
realized by de Wit and Freedman \cite{dWF77}. They remarked the presence of complicated, non-polynomial
structures in the scalars and pseudo-scalars of the theory. After that, Cremmer and Julia described in
detail the component structure of this theory \cite{CJ78}, \cite{CJ79}, obtaining it by dimensional
reduction of the 11 dimensional supergravity theory to 4 dimensions. In particular, they established that
the scalars take part of a non-linear $\s$ model and they live on the $E_7/SU(8)$ coset space. De Wit and
Nicolai also resumed the properties of this theory \cite{dWN82} with particular attention on separating
deducible properties and conjectures which yield the known and accepted structure.

The well-known methods of superspace geometry, as developed by Wess and Zumino \cite{WZ77}  and resumed in \cite{WB83} were generalized to extended supergravity and
were applied to the $N=8$ case \cite{BH79}, \cite{HL81}. In these papers, the authors used
ordinary extended superspace, without any additional bosonic coordinates. In this approach the
identification of both the vectors and the scalars can be done through their field strengths in torsion components,
nevertheless, for more clearness, the necessary number of gauge vectors is introduced explicitly in the covariant
derivatives. In order to render the description more transparent, Howe and Lindstr\"om in the appendix of 
\cite{HL81} and Siegel \cite{Sie81b} introduce additional bosonic coordinates in superspace. 
This second approach is completely equivalent with the former one, though, it gives a more 
natural interpretation of graviphotons: they are identified as the gauge vectors corresponding to local translations
in the direction of the additional bosonic coordinates. Concerning the scalars, they are identified in a non-trivial way in some components of the
super-vielbein in the extra bosonic sector. In reference \cite{How82} a detailed presentation is given of this
second approach.

The aim of this paper is to present an approach to the identification of the $N=8$ supergravity in
central charge superspace, generalizing the constructions 
applied to $N=2$ \cite{AGHH99} and $N=4$ \cite{GHK01}. Here, the vectors are identified in the frame
components with central charge flat index and the corresponding gauge transformations are realized as
superspace diffeomorphisms in the central charge directions. Moreover, we show that it is possible to
identify the scalars of the theory directly, in torsion components of 0 canonical dimension. Also, the
three silver rules for maximally dualised extended supergravities \cite{Jul98} are recovered as
consequences of natural and simple constraints on the geometry. Namely, we find that the moduli space of
the scalar fields has a $E_{7(+7)}/SU(8)$ coset space structure, that these scalar fields serve as
converters between $SU(8)$ indices and central charge indices (corresponding to $E_{7(+7)}$
representation indices), and also, that the vectors taking part of the multiplet are self--dual or
anti--self--dual in the $SU(8)$ basis while they satisfy a twisted self--duality relation in the central
charge basis.

The paper is organized as follows. In section 2, we review the basic structure of central charge
superspace. Section 3 is devoted to the identification of the multiplet: physical fields are identified
in the vielbein and in the torsion, and their main properties are deduced from the geometric structure.
Then, we compute the supergravity transformations of the fields in section 4. We finish with the
equations of motion which are presented in section 5.

%%%%%%%%%%%%%%%%%%%%%%%%%%%%%%%%%%%%%%%%%%%%%%%%%%%
\section{General geometric structure}

%%%%%%%%%%%%%%%%%%%%%%%%%%%%%%%%%%%%%%%%%%%%%%%%%%%

The geometrical description of extended supergravity theories is based on a 
generalization to the extended case of standard $N=1$ superspace methods \cite{WB83}, \cite{GGRS83}. 
The Abelian gauge vectors which appear for higher $N$ have a natural interpretation
in central charge superspace: they appear as components of the superspace frame. As
such, their gauge transformations appear on the same footing as
space-time diffeomorphisms and local supersymmetry transformations. General ideas about supergravity in
central charge superspace are presented in detail in \cite{AGHH99} and \cite{GHK01}. Nevertheless, we
recall here briefly the basic notions we use throughout the article.

%==========================================
%\subsection{Basic superspace structures}
%==========================================

Consider the central charge superspace with frame $E^\ca=(E^a,E_\ta^\a,E^\ta_\da, E^\au)$, where $a$,
$\a$, $\da$ denote the usual vector and Weyl spinor indices, capital indices $\ta$ count the number of
supercharges and boldface indices $\au$ the number of central charges. The structure group, which is
chosen to be $SL(2,C)\otimes SU(8)$, acts on the frame, $\d_X E^\ca=E^\cb X_\cb{}^\ca$, and the
corresponding covariant derivatives are defined in terms of the connection 1--form $\F_\cb{}^\ca$. The
representation of the structure group on flat indices is block-diagonal with respect to the space-time,
spinorial and central charge sector. The a priori non-zero connection components $\F_\cb{}^\ca$ are then
\be
\F_b{}^a\,,
\qquad\F^\tb_\b{}^\a_\ta\ =\
\d^\tb_\ta\F_\b{}^\a+\d_\b^\a\F^\tb{}_\ta\,,
\qquad\F_\tb^\db{}_\da^\ta\ =\
\d_\tb^\ta\F^\db{}_\da+\d^\db_\da\bar{\F}
_\tb{}^\ta\,,\qquad\F_\au{}^\az\,,
\ee
with $\F^\ta{}_\ta=0$ and $\F^\tb{}_\ta + \bar{\F}_\ta{}^\tb = 0$, as properties of the vector
representation of the Lie algebra $su(8)$ \footnote{We note $SU(8)$ the Lie group and
$su(8)$ its Lie algebra}. (Conventions and notations concerning vector and spinor
representation of the Lorentz group are those of \cite{BGG01}, while some useful properties of $SU(N)$
tensors are given in appendix \ref{deltas}.)

The torsion and the curvature are given as differential forms in central charge 
superspace 
\bea
T^\ca&=&DE^\ca\ =\ \vd E^\ca+E^\cb\F_\cb{}^\ca\,,\label{torsion}\\[2mm]
R_\cb{}^\ca&=&\vd\F_\cb{}^\ca+\F_\cb{}^\cc\F_\cc{}^\ca\,,
\eea
and satisfy the Bianchi identities
\bea
DT^\ca&=&E^\cb R_\cb{}^\ca\,,\label{idbT}\\[2mm]
DR_\cb{}^\ca&=&0\,,\label{idbR}
\eea
where the Bianchi identity of the torsion \equ{idbT} is a consequence of the action of two covariant
derivatives on a covariant vector $u^\ca$,
\be
DDu^\ca\ =\ u^\cb R_\cb{}^\ca\,,
\eqn{algebra}
and the definition of the torsion \equ{torsion}.

The supergravity transformations (or Wess-Zumino transformations) are defined as usual \cite{WB83},
\cite{BGG01} to be a special combination of diffeomorphisms and gauge transformations such that they
transform covariant vectors into covariant vectors:
\be
\dwz_\x\doteq (L_\x + \d_X)\loco_{X=\imath_\x \F}\,,
\ee
acting on covariant vectors by the ``covariant Lie derivative" $\cl_\x=\imath_\x D+D \imath_\x$. In
particular, the supergravity transformations of the vielbein and the connection
are the following:
\bea
\dwz_\x E^\ca&=&D\x^\ca+\imath_\x T^\ca\,,\label{dframe}\\[2mm]
\dwz_\x\F_\cb{}^\ca&=&\imath_\x R_\cb{}^\ca\,.
\eea

The geometrical description of the $N=8$ supergravity theory in this context, is to identify first the
components of the multiplet in the geometrical objects reviewed above (frame, torsion, connection,
curvature) in such a way that they transform under supergravity transformations between themselves. Once
the identification of the component fields is done and in order to convince oneself that the identified
theory is nothing else but the $N=8$ supergravity described in the original works on the component level,
one has to compare the supersymmetry transformations as well as the equations of motion we can deduce
from the geometrical description with those given in the component formalism \cite{CJ78}, \cite{dWN82}.

This is the aim and the strategy of the next sections.

%%%%%%%%%%%%%%%%%%%%%%%%%%%%%%%%%%%%%%%%%%%%%%%%%%%%%%%%%%%%%%%%%%%%%%%%%%%%%%%%%%%%
\section{Identification and properties of the multiplet}
%%%%%%%%%%%%%%%%%%%%%%%%%%%%%%%%%%%%%%%%%%%%%%%%%%%%%%%%%%%%%%%%%%%%%%%%%%%%%%%%%%%%
\label{id}

We will identify in this section the component fields (one graviton, 8
Rarita-Schwinger fields, 28 graviphotons, 56 helicity 1/2 fields and 70 scalars) 
in the context of the basic geometrical objects of central charge superspace.

%=============================================================================
\subsection{Identification of the gauge component fields in the super-vielbein}
%=============================================================================

In analogy with general relativity, where the graviton is identified in the vierbein, 
central charge superspace provides a unified geometric interpretation of the graviton,
gravitini and graviphotons in the frame $E^\ca$,
\be
E^a\doubar\ =\ \vd x^m e_m{}^a~,\quad E_\ta^\a\doubar\ =\ \frac{1}{2}\,\vd x^m
\p_m{}_\ta^\a~,
\quad E^\ta_\da\doubar\ =\ \frac{1}{2}\,\vd x^m \bar{\p}_m{}^\ta_\da~,
\quad E^\au\doubar\ =\ \vd x^m v_m{}^\au~,
\eqn{frame}
where the double bar projects at the same time on the vector coefficient of the differential form and on
the lowest superfield component \cite{BBG87}.

Also, there are a priori other independent gauge fields in the theory, namely, the connection fields.
Concerning the connection in the central charge sector, let us adopt in the following the requirement
that the representation of the structure group on central charge indices be trivial,
\be
\F_\au{}^\az\ =\ 0\,.
\eqn{fiuz}
Concerning the other components, the double bar projections of the connection one--forms
\be
\F_b{}^a\doubar\ =\ \vd x^m\F_m{}_b{}^a~,\qquad\F^\tb{}_\ta\doubar\
=\ \vd x^m\F_m{}^\tb{}_\ta
\ee
give the ordinary Lorentz and $SU(8)$ connections. However, as it will be shown in the next sections (see
equations \equ{fimba} and \equ{fim_su8}), once the constraint \equ{fiuz} adopted, in the case of the
on--shell $N=8$ supergravity both the Lorentz and the $SU(8)$ connections are given as functions of the
other component fields and their space-time derivatives.

As the remaining component fields of the multiplet, the scalars and the helicity 1/2 fields, have to
complete the above gauge fields into a supersymmetry multiplet. We are looking for them in the
supergravity transform of the frame \equ{dframe}, that is in torsion components, which satisfy the Bianchi
identity \equ{idbT}, or displayed on 3--form coefficients,
\be
\left(_{\cd\cc\cb}{}^\ca\right)_T\quad:\quad E^\cb E^\cc E^\cd
\left(\cd_\cd
T_{\cc\cb}{}^\ca+T_{\cd\cc}{}^\cf T_{\cf\cb}{}^\ca-R_{\cd\cc\cb}{}^\ca
\right)=0\,.
\eqn{idbT_comp}
Torsion components also appear in the algebra of covariant derivatives \equ{algebra}
\be
(\cd_\cc,\cd_\cb)u^\ca\ =\ -T_\cc{}_\cb{}^\cf \cd_\cf u^\ca+R_\cc{}_\cb{}_\cf{}^\ca u^\cf\,,
\eqn{algebra_comp}
when it is acting on a covariant vector $u^\ca$.

Recall, that in general relativity the Bianchi identities \equ{idbT} and \equ{idbR} are independent,
while for supergravity defined in ordinary extended superspace without central charge coordinates, 
Dragon's theorem \cite{Dra79} tells us that the Bianchi identities for the curvature
\equ{idbR} are a consequence of the Bianchi identities for the torsion \equ{idbT}. In this latter
case, the torsion is considered as the fundamental object of the geometry and all curvature components
are expressed as functions of the torsion components and their covariant derivatives. However, as the
theorem of Dragon is based on the relation between the representations of the structure group on bosonic
and fermionic indices, in the presence of central charge coordinates in general it ceases to be valid as
it stands \cite{Kis00}. Nevertheless, assuming trivial gauge structure \equ{fiuz} in
the central charge sector, Dragon's theroem remains valid and it is sufficiant to
investigate the first set \equ{idbT_comp} of Bianchi identities.

The constraints 
\be
T^\tc_\g{}_b{}^a\ =\ 0\,,\qquad T_\tc^\dg{}_b{}^a\ =\ 0\,,\qquad T_{cb}{}^a\ =\ 0\,,
\eqn{conventional}
are the usual conventional ones \cite{Mul86b}, \cite{Kis00},
which are nothing else but some redefinitions of the supervielbein and of the Lorentz connection. 
Other less conventional constraints, as for example 
\be
T_\az{}_\cb{}^\ca\ =\ 0\,,
\eqn{Tzba}
serve to reduce the number of independant fields and may also imply equations
of motion.

The remaining constraints, at canonical dimension 0 and 1/2, allow the
identification of the scalars and the 1/2 helicity fields.
They will be presented in the next two paragraphes in some more detail. 
Once the constraints on torsion components are imposed, their consistency 
with Bianchi identities \equ{idbT_comp} must be checked, which has been
carefully done.

%===================================================================================
\subsection{Constraints at dimension 0, identification of the scalars}
%===================================================================================

The geometrical description of the $N=8$ supergravity multiplet in central charge superspace is based on a
set of natural constraints at canonical dimension 0. This type of constraints was already used to identify
the $N=2$ minimal supergravity multiplet \cite{AGHH99} as well as to identify the $N=4$ supergravity with
antisymmetric tensor \cite{GHK01}, \cite{GK02}, the Nicolai--Townsend multiplet.
They remain the same in the present case of $N=8$ supergravity :
\be
 T{^\tc_\g}{^\tb_\b}{^a_{}}\ =\ 0~, \qquad
 T{^\tc_\g}{^\db_\tb}{^a}\ =\ -2\vi\d{^\tc_\tb}(\s{^a}\eps){_\g}{^\db}~, \qquad
 T{^\dg_\tc}{^\db_\tb}{^a}\ =\ 0~,
\eqn{T01}
\be
T{^\tc_\g}{^\tb_\b}{^\au}\ =\ \eps_{\g\b}T^{[\tc\tb]\au}~,\qquad T^\tc_\g{}^\db_\tb{}^\au\ =\ 0~, \qquad
T{^\dg_\tc}{^\db_\tb}{^\au}\ =\
\eps^{\dg\db}T{_{[\tc\tb]}}{}^\au~.
\eqn{T02}

As in the $N=4$ case, we expect that the objects $T^{[\tc\tb]\au}$ and $T{_{[\tc\tb]}}{}^\au$, which can
be organized in matrix form as
\be
T\doteq\left(
\begin{array}{c}T^{[\td\tc]}{}^\au\\T_{[\td\tc]}{}^\au\end{array}\right)\,,
\ee
play an important role in the identification of the scalars. One of the most important assumptions we
make in order to identify the $N=8$ supergravity multiplet is to suppose that there exists a matrix $S$
with components
\be
S\doteq\left(
\begin{array}{cc}S_\au{}_{[\td\tc]}&S_\au{}^{[\td\tc]}\end{array}\right)\,,
\ee
such that the components of $T$ and $S$ satisfy
\bea
\left(\begin{array}{cc}T^{[\td\tc]}{}^\au S_\au{}_{[\tb\ta]}&
T^{[\td\tc]}{}^\au S_\au{}^{[\td\tc]}\\
T_{[\td\tc]}{}^\au S_\au{}_{[\tb\ta]}& T_{[\td\tc]}{}^\au S_\au{}^{[\td\tc]}
\end{array}\right)&=&
\left(\begin{array}{cc}\frac{1}{2}\d^{\td\tc}_{\tb\ta}&0\\
0&\frac{1}{2}\d_{\td\tc}^{\tb\ta}
\end{array}\right)\label{id1}\\[3mm]
S_\au{}_{[\tb\ta]}T^{[\tb\ta]}{}^\az+S_\au{}^{[\tb\ta]}T_{[\tb\ta]}{}^\az &=&\d^\az_\au\,.\label{id2}
\eea

Recall, that we did not fix a priori the number of central charge indices $\au$. However, the assumptions
above imply that the matrices $T$ and $S$ are square matrices of dimension $N(N-1)$ and of maximal
rank\footnote{The demonstration is immediate using the property $rank(AB)\leq min(rank(A),\,rank(B))$ for
all matrices $A$ and $B$ with dimensions such that the product of them is well-defined.}. So, these
assumptions fix the number of central charge indices $\au$ to be $N(N-1)=56$, and we can write the relations
\equ{id1} and \equ{id2} as
\be
TS=\mathbf{1}_{56}\,,\qquad ST=\mathbf{1}_{56}\,.
\ee
In this sense $S$ is the inverse of the matrix $T$, constructed from the Lorentz scalars contained
in the torsion components of zero canonical dimension \equ{T02}.

As a matter of fact, these matrices serve as converters between the central charge basis (indices \au) and
the $SU(8)$ basis in the antisymmetric representation (indices ${}_{[\td\tc]}$ and ${}^{[\td\tc]}$). For
instance all object $X^\au$ can be converted in the $SU(8)$ basis using the matrix $S$:
\be
\left(
\begin{array}{cc}X_{[\td\tc]}&X^{[\td\tc]}\end{array}
\right)\ =\
X^\au\left(
\begin{array}{cc}S_\au{}_{[\td\tc]}&S_\au{}^{[\td\tc]}\end{array}
\right).
\eqn{XDC}
Inversely, one can come back to the central charge basis using the matrix $T$:
\be
X^\au\ =\ \left(
\begin{array}{cc}X_{[\td\tc]}&X^{[\td\tc]}\end{array}
\right)\left(
\begin{array}{c}T^{[\td\tc]}{}^\au\\T_{[\td\tc]}{}^\au\end{array}
\right)\ =\
X_{[\td\tc]}T^{[\td\tc]}{}^\au+X^{[\td\tc]}T_{[\td\tc]}{}^\au.
\ee

We identify the scalars of the multiplet as the lowest superfield components of the scalar superfields
$T$:
\be
T\loco\ =\ \ct\,,\qquad S\loco\ =\ \cs\,.
\ee
However, at this stage we have a problem with the degrees of freedom contained in the objects $\Sc$.
Namely, in the $N=8$ supergravity multiplet we expect to have 70 scalars, while the matrix $T$ identified
in torsion components, and therefore its lowest superfield components $\ct$, have a priori $56\times 56$
independent components. This problem will be solved in paragraphe \ref{scalaire}, at dimension 1,
where we will rather count the degrees of freedom in the field strength $(\cd_m\ct)\cs$.

%===================================================================================
\subsection{Constraints at dimension 1/2, identification of the 1/2 helicity fields}
%===================================================================================

In turn, the helicity 1/2 fields, or gravigini fields, are identified as usual \cite{How82}, \cite{GG83}
in the dimension 1/2 torsion component
\be
T^\tc_\g{}^\tb_\b{}^\ta_\da\ =\ \eps_{\b\g}T^{[\tc\tb\ta]}{}_\da\,,\qquad
T_\tc^\dg{}_\tb^\db{}_\ta^\a\ =\ \eps^{\db\dg}T_{[\tc\tb\ta]}{}^\a\,,
\eqn{spineurs}
as the lowest superfield component of the gravigini superfields $T_{[\tc\tb\ta]}{}^\a$ and
$T^{[\tc\tb\ta]}{}_\da$,
\be
T_{[\tc\tb\ta]}{}^\a\loco\ =\ \T_{[\tc\tb\ta]}{}^\a\,,
\qquad T^{[\tc\tb\ta]}{}_\da\loco\ =\ \T^{[\tc\tb\ta]}{}_\da\,.
\ee

Recall that our aim is to describe the $N=8$ on--shell supergravity multiplet containing no other
independent helicity 1/2 fields as these $C^3_8=56$ ones in the totally antisymmetric representation.
Therefore, the constraints at canonical dimension 1/2 are chosen in such a way that on one hand they eliminate all
dimension 1/2 fields which are not in the good representation of $SU(8)$, and, on the other hand all the remaining 1/2
helicity fields are linear combinations of $T_{[\tc\tb\ta]}{}_\a$ and $T^{[\tc\tb\ta]}{}^\da$.

More precisely, we require first 
\be
T^\tc_\g{}^\tb_\b{}_{\ta}^\a\ =\ 0\,,\qquad\quad T_\tc^\dg{}_\tb^\db{}^{\ta}_\da\ =\ 0\,,\qquad\quad
T^\tc_\g{}^\db_\tb{}_\da^\ta\ =\ 0\,,\qquad\quad T_\tc^\g{}_\db^\tb{}^\da_\ta\ =\ 0\,.
\ee

The case of torsion components of the form $T_{\cc\cd}{}^\au$ deserves more attention.
Recall that we have already eliminated those with at least one central charge differential form 
index. As to those with a structure group central charge index, the possibility to 
pass from the central charge basis (indices $\au$) to the $SU(8)$ basis
(antisymmetric combination $[BA]$) will be crucial for the following discussion.
Candidates of torsion components which allow to accomodate $T_{[\tc\tb\ta]}{}_\a$
or $T^{[\tc\tb\ta]}{}^\da$ are 
\be
X_\d^{[\td\tc\tb]\au}\ =\ \cd^\td_\d T^{[\tc\tb]\au}\,,
\qquad\quad X^\dd_{[\td\tc\tb]}{}^\au\ =\ \cd_\td^\dd T_{[\tc\tb]}{}^\au\,,
\ee
and $T_{\a\ta}{}^\au$, $T^{\da\ta}{}^\au$, which appear in the expression of the torsion components
\be
T^\tc_\g{}_b{}^\au\ =\ -2\vi(\s_b)_{\g\dg}T^{\dg\tc}{}^\au\,,
\qquad\qquad T_\tc^\dg{}_b{}^\au\ =\ -2\vi(\bs_b)^{\dg\g}T_{\g\tc}{}^\au\,.
\ee
As a consequance, we require 
\bea
T_{\a\tc}{}^\au
\left(
\begin{array}{cc}S_\au{}_{[\tb\ta]}&S_\au{}^{[\tb\ta]}\end{array}
\right)& =&\left(
\begin{array}{cc}\a T_{\g[\tc\tb\ta]}&0\end{array}
\right)
\\[2mm]
T^{\da\tc}{}^\au
\left(
\begin{array}{cc}S_\au{}_{[\tb\ta]}&S_\au{}^{[\tb\ta]}\end{array}
\right)& =&\left(
\begin{array}{cc}0&\bar{\a} T^{\dg[\tc\tb\ta]}\end{array}
\right)
\\[2mm]
X_\g^{[\tf\td\tc]}{}^\au
\left(
\begin{array}{cc}S_\au{}_{[\tb\ta]}&S_\au{}^{[\tb\ta]}\end{array}
\right)&=&\left(
\begin{array}{cc}0&\b\e^{\tf\td\tc\tb\ta\tg_1\tg_2\tg_3} T_{\g[\tg_1\tg_2\tg_3]}\end{array}
\right)
\\[2mm]
X^\dg_{[\tf\td\tc]}{}^\au
\left(
\begin{array}{cc}S_\au{}_{[\tb\ta]}&S_\au{}^{[\tb\ta]}\end{array}
\right)& =&\left(
\begin{array}{cc}\bar{\b}\e_{\tf\td\tc\tb\ta\tg_1\tg_2\tg_3} T^{\dg[\tg_1\tg_2\tg_3]}&0\end{array}
\right)\,.
\eea

These assumptions assure that the gravigini fields $T_{[\tc\tb\ta]}{}_\a$ and $T^{[\tc\tb\ta]}{}^\da$ are
the only 1/2 helicity fields in the geometry, and in particular, all the torsion components of canonical
dimension 1/2 can be expressed using these fields (see appendix \ref{sol}). Also, with these additional
assumptions the Bianchi identities $\brg{^\td_\d{}^\tc_\g{}^\tb_\b{}^\au}$,
$\brg{^\td_\d{}^\tc_\g{}_\tb^\db{}^\au}$ with their complex conjugates imply that the Lorentz scalar $T$
transforms under supersymmetry transformations into the helicity 1/2 fields:
\bea
(\cd^\td_\d T)S&=&\left(
\begin{array}{cc}
0&\b\e^{\tb\ta\tb'\ta'\td\te\tf\tg} T_{[\te\tf\tg]\d}\\
-\frac{1}{2}\frac{1}{3!}
\d^{\td\te\tf\tg}_{\tb\ta\tb'\ta'}T_{[\te\tf\tg]\d}
-\frac{(16\a-1)}{2}\d^{\td\tc}_{\tb\ta}T_{[\tc\tb'\ta']\d}&0
\end{array}\right)\label{D1spinTS}
\\[2mm]
(\cd_\td^\dd T)S&=&\left(
\begin{array}{cc}
0&-\frac{1}{2}\frac{1}{3!}
\d_{\td\te\tf\tg}^{\tb\ta\tb'\ta'}T^{[\te\tf\tg]\dd}
-\frac{(16\bar{\a}-1)}{2}\d_{\td\tc}^{\tb\ta}T^{[\tc\tb'\ta']\dd}\\
\bar{\b}\e_{\tb\ta\tb'\ta'\td\te\tf\tg} T^{[\te\tf\tg]\dd}&0
\end{array}\right)\label{D2spinTS}
\eea
with $\a$, $\bar{\a}$, $\b$ and $\bar{\b}$ some complex parameters, which will be determined by
consistency requirements at higher canonical dimensions, in sections \ref{selfdphoton} and \ref{scalaire}.

%===================================================================================
\subsection{Self-duality and anti-self-duality of the graviphotons}
%===================================================================================
\label{selfdphoton}

The use of central charge superspace allowed to identify the gauge vectors of the multiplet
in the super-vielbein \equ{frame}. Notice, on the one hand that at this stage we identified in the
geometry as many gauge vectors $v_m{}^\au$ as the number of central charge indices, that is $56$. On the
other hand, the number of graviphotons taking part of the $N=8$ supergravity multiplet is only the half of
that, that is $28$. The aim of this paragraph is to clarify this problem by analyzing how the Bianchi
identities imply specific properties satisfied by the field strength of the gauge vectors $v_m{}^\au$,
reducing their degrees of freedom to the half.

Since the vectors are identified in the vielbein, their super-covariant field strength is the torsion
component $T_{ba}{}^\au$, which we denote in the following by $F_{ba}{}^\au$ and which in turn can be
converted in the $SU(8)$ basis by
\be
\left(
\begin{array}{cc}F_{ba}{}_{[\td\tc]}&F_{ba}{}^{[\td\tc]}\end{array}
\right)\ =\
F_{ba}{}^\au\left(
\begin{array}{cc}S_\au{}_{[\td\tc]}&S_\au{}^{[\td\tc]}\end{array}
\right).
\eqn{FDC}
It is worthwhile to present here briefly the intriguing interplay of the Bianchi identities in order to
determine the properties of this object.

First of all, recall the very general result of the Bianchi identities
$\brg{^\dd_\td{}^\dg_\tc{}^\tb_\b{}^\a_\ta}$ and $\brg{_\d^\td{}_\g^\tc{}_\tb^\db{}_\da^\ta}$ concerning
the spinorial derivatives of the 1/2 helicity fields
\bea
\cd^\td_{(\b} T_{\a)[\tc\tb\ta]}
=-\vi\d_{\tc\tb\ta}^{\td\te\tf}G_{(\b\a)[\te\tf]}\,,\\[2mm]
\cd_\td^{(\db} T^{\da)[\tc\tb\ta]}
=-\vi\d^{\tc\tb\ta}_{\td\te\tf}G^{(\db\da)[\te\tf]}\,.
\eea
Here the superfields $G_{(\b\a)[\tb\ta]}$ and $G^{(\db\da)[\tb\ta]}$ are the self--dual and respectively,
the anti--self--dual part of the a priori arbitrary antisymmetric tensor superfields $G_{ba}{}_{[\tb\ta]}$
and $G_{ba}{}^{[\tb\ta]}$, appearing in the torsion components
\bea
T_\tc^\dg{}_b{}_\ta^\a&=&\frac{1}{2}(\bs^f)^{\dg\a}
\left(\eta_{fb}T_{(\tc\ta)}-G_{fb}{}_{[\tc\ta]}\right)\,,\\[2mm]
T^\tc_\g{}_b{}^\ta_\da&=&\frac{1}{2}(\s^f)_{\g\da}
\left(\eta_{fb}T^{(\tc\ta)}-G_{fb}{}^{[\tc\ta]}\right)\,,
\eea
as they are given by the Bianchi identities $\brg{_\td^\dd{}_\tc^\dg{}_b{}^a}$ and
$\brg{^\td_\d{}^\tc_\g{}_b{}^a}$. Now the question is how to relate the antisymmetric tensors
$G_{ba}{}_{[\tb\ta]}$ and $G_{ba}{}^{[\tb\ta]}$ to the field strength $F_{ba}{}^\au$ of the graviphotons,
since a relation between these two objects would insure that, as expected, the supersymmetry transform of
the 1/2 helicity fields contains the field strength of the gauge vectors.

The response is given by the Bianchi identity $\brg{^\td_\d{}^\dg_\tc{}_b{}^\au}$, which turns out to be
a real mine of information. Namely, it implies that the antisymmetric tensors $G_{ba}{}_{[\tb\ta]}$ and
$G_{ba}{}^{[\tb\ta]}$ are related to the $SU(8)$ components of the field strength of the graviphotons in
a very simple way:
\be
G_{ba}{}_{[\tb\ta]}\ =\ -8\vi F_{ba}{}_{[\tb\ta]}\,,
\qquad\qquad G_{ba}{}^{[\tb\ta]}\ =\ -8\vi F_{ba}{}^{[\tb\ta]}\,.
\ee
Moreover, it implies that the parts of the field strength which are not present in the spinorial
derivative of the helicity 1/2 fields vanish in the linear approach. They are given as quadratic terms in
the gravigini:
\bea
F^{(\dd\db)}{}_{[\tb\ta]}&=&\frac{\bar{\a}\bar{\b}}{3!}\,
\e_{\tb\ta\tf_1...\tf_6}T^{[\tf_1\tf_2\tf_3]\dd}T^{[\tf_4\tf_5\tf_6]\db}\,,\label{f1}\\[2mm]
F_{(\d\b)}{}^{[\tb\ta]}&=&\frac{\a\b}{3!}\,
\e^{\tb\ta\tf_1...\tf_6}T_{[\tf_1\tf_2\tf_3]\d}T_{[\tf_4\tf_5\tf_6]\b}\,.\label{f2}
\eea
These are the relations which permit us to write down the self--duality properties of the graviphotons
\equ{self} and explain the reduction of the graviphotons' degrees of freedom.

Finally, the same Bianchi identity implies the vanishing of the superfields $T^{(\tb\ta)}$ and
$T_{(\tb\ta)}$, gives the antisymmetric part of the spinorial derivative of the gravigini as a quadratic
term in themselves
\bea
\cd^\td_{[\b} T_{\a][\tc\tb\ta]}
&=&-\frac{\bar{\b}}{2}\,\eps_{\b\a}\,
\e_{\tc\tb\ta\te\tf\tg\th\ti}T_\da^{[\td\te\tf]}T^{[\tg\th\ti]\da}\,,\\[2mm]
\cd_\td^{[\db} T^{\da][\tc\tb\ta]}
&=&-\frac{\b}{2}\,\eps^{\db\da}\,
\e^{\tc\tb\ta\te\tf\tg\th\ti}T^\a_{[\td\te\tf]}T_{[\tg\th\ti]\a}\,,
\eea
and fixes the parameters $\a$ and $\bar{\a}$ by implying the relations
\be
\left(\a-\frac{1}{16}\right)F_{(\d\b)}{}_{[\tb\ta]}\ =\ 0\,,
\qquad\qquad\left(\bar{\a}-\frac{1}{16}\right)F^{(\dd\db)}{}^{[\tb\ta]}\ =\ 0\,.
\eqn{alpha}
Recall, that the parts of the field strength $F$ which appear in relations \equ{alpha} are those, in
which the gravigini fields transform under supersymmetry and we require that they do not vanish. So, the
parameters $\a$ and $\bar{\a}$ are determined to be $\a = \bar{\a} = 1/16$.

As a conclusion of this paragraph let us denote the objects
\bea
\fp_{ba}{}_{[\tb\ta]}&=&F_{ba}{}_{[\tb\ta]}-
\frac{1}{16}\frac{\bar{\b}}{3!}\e_{\tb\ta\tf_1...\tf_6}\,T^{[\tf_1\tf_2\tf_3]}\bs_{ba}T^{[\tf_4\tf_5\tf_6]}
\label{fp}\\[2mm]
\fm_{ba}{}^{[\tb\ta]}&=&F_{ba}{}^{[\tb\ta]}-
\frac{1}{16}\frac{\b}{3!}\e^{\tb\ta\tf_1...\tf_6}\,T_{[\tf_1\tf_2\tf_3]}\s_{ba}T_{[\tf_4\tf_5\tf_6]}
\label{fm}
\eea
which in virtue of the relations \equ{f1} and \equ{f2} satisfy the self--duality and anti--self--duality
relations
\be
\frac{\vi}{2}\e^{dcba}\fp_{ba}{}_{[\tb\ta]}\ =\ \fp{}^{dc}{}_{[\tb\ta]}\,,\qquad
\frac{\vi}{2}\e^{dcba}\fm_{ba}{}^{[\tb\ta]}\ =\ -\fm{}^{dc}{}^{[\tb\ta]}\,,
\eqn{self}
and are the parts of the graviphoton field strength which effectively take part of the multiplet. These
self--duality relations can also be given in the central charge basis as
\bea
\frac{\vi}{2}\e^{dcba}F_{ba}{}^\au&=&F^{dc}{}^\az\left(S\om T\right)_\az{}^\au\nn\\[2mm]
&&-\frac{1}{8}\frac{\bar{\b}}{3!}
\e_{\tb\ta\tf_1...\tf_6}(T^{[\tf_1\tf_2\tf_3]}\bs_{ba}T^{[\tf_4\tf_5\tf_6]})T^{[\tb\ta]}{}^\au\nn\\[2mm]
&&+\frac{1}{8}\frac{\b}{3!}
\e^{\tb\ta\tf_1...\tf_6}(T_{[\tf_1\tf_2\tf_3]}\s_{ba}T_{[\tf_4\tf_5\tf_6]})T_{[\tb\ta]}{}^\au\,,
\label{TS_nonl}
\eea
with
\be
\om\ =\ \left(
\begin{array}{cc}
\frac{1}{2}\d^{\td\tc}_{\tb\ta}& 0\\
0&-\frac{1}{2}\d_{\td\tc}^{\tb\ta}
\end{array}\right)
\ee
an $56\times 56$ matrix. In the linear approach one can recognize in \equ{TS_nonl} the twisted
self--duality relation for graviphotons \cite{Jul98}, which is also called the third silver rule of
supergravity%
\footnote{It is possible to construct an object $\tilde{F}_{ba}{}^\au$ satisfying the pure twisted
self--duality relation
\[
\frac{\vi}{2}\e^{dcba}\tilde{F}_{ba}{}^\au\ =\ \tilde{F}^{dc}{}^\az\left(S\om T\right)_\az{}^\au
\]
even in the full non--linear case, using the self--dual and anti--self--dual field strengths
$\fp_{ba}{}_{[\tb\ta]}$ and $\fm_{ba}{}^{[\tb\ta]}$:
\[
\tilde{F}_{ba}{}^\au\ =\ \left(
\begin{array}{cc}\fp_{ba}{}_{[\td\tc]}&\fm_{ba}{}^{[\td\tc]}\end{array}
\right)\left(
\begin{array}{c}T^{[\td\tc]}{}^\au\\T_{[\td\tc]}{}^\au\end{array}
\right).
\]}. Let us anticipate here, that the matrix $\om$ plays the r\^ole of an invariant operator acting on the
$56$ dimensional representation of the Lie group $K$ for a supergravity theory, where the scalars are
organized in a $G/K$ non--linear sigma model. In our case one finds a $E_{7(+7)}/SU(8)$ non--linear sigma
model structure for the scalars
-- but this will be the subject of the next paragraph.

%===================================================================================
\subsection{The $E_{7(+7)}/SU(8)$ non--linear sigma model}
%===================================================================================
\label{scalaire}

As already observed on the component level \cite{CJ79}, \cite{dWN82} and put in evidence in the former
superspace approaches \cite{BH79}, \cite{HL81}, \cite{How82}, the hypothesis that the scalars take part
of an $E_{7(+7)}/SU(8)$ non--linear sigma model is compatible with the structure of the multiplet.
General features of $G/K$ non--linear sigma models with $G$ a non--compact Lie group and $K$ its maximal
compact sub--group can be found in \cite{CWZ69}, \cite{CCWZ69}, \cite{GZ81}.

Recall however, that in component approaches this structure is related to the existence of a duality
invariance \cite{CJ79}, \cite{GZ81}, while the identification of $E_{7(+7)}$ as the duality group is based
only on considerations on the dimensions of the interplaying Lie groups \cite{CJ79}, \cite{dWN82}. Then,
the existing geometrical descriptions, in order to recover this structure, use constraints inspired by
the $56$ dimensional, fundamental representation of the Lie algebra of $E_{7(+7)}$ on the supercovariant
field strength of the scalars. The aim of this paragraph is to analyze the properties concerning these
aspects implied naturally by the constraints presented so far.

As in the case of the gauge vectors there is an ensemble of Bianchi identities which interplay in order
to give the properties of the field strength of the scalars identified in the $56\times 56$ matrix
$(\cd_a T)S$.

First of all, as we expect to identify the field strength of the scalars in the supersymmetry transform of
the gravigini fields, let us recall the general result of the analysis of Bianchi identities
$\brg{^\td_\d{}^\tc_\g{}^\tb_\b{}^\ta_\da}$ and $\brg{^\dd_\td{}^\dg_\tc{}^\db_\tb{}^\a_\ta}$. Namely,
that they are satisfied if and only if the covariant spinorial derivatives of the spinor fields
$T_{[\tc\tb\ta]\a}$ and $T^{[\tc\tb\ta]\da}$ giving a Lorentz vector are totally antisymmetric in their
$SU(8)$ indices:
\be
\cd^\td_\d T^{[\tc\tb\ta]\da}\ =\ P_\d{}^\da{}^{[\td\tc\tb\ta]}\,,\qquad\qquad
\cd_\td^\dd T_{[\tc\tb\ta]\a}\ =\ P_\a{}^\dd{}_{[\td\tc\tb\ta]}\,,
\ee
with $P_\b{}^\db{}^{[\td\tc\tb\ta]}$ and $P_\b{}^\db{}_{[\td\tc\tb\ta]}$ a priori some arbitrary
superfields.

Now again, the essential question is how can we relate the field strength of the scalars, $(\cd_b T)S$, to
the superfields $P_b{}^{[\td\tc\tb\ta]}$ and $P_b{}_{[\td\tc\tb\ta]}$. In order to answer this question,
notice that the covariant derivatives $\cd_bT^{[\td\tc]}{}^\au$ and $\cd_bT_{[\td\tc]}{}^\au$ appear
explicitly in the Bianchi identities $\brg{_\d^\td{}_\g^\tc{}_b{}^\au}$ and
$\brg{_\td^\dd{}_\tc^\dg{}_b{}^\au}$. These are the identities which imply that the field strength $(\cd_b
T)S$ of the scalars should take the form
\be
(\cd_b T)S \ =\ \left(
\begin{array}{cc}
-2\d^{[\td}_{[\tb}\hat{T}_b{}^{\tc]}{}_{\ta]}& -\frac{\vi}{4}P_b{}^{[\td\tc\tb\ta]}\\
-\frac{\vi}{4}P_b{}_{[\td\tc\tb\ta]}&2\d^{[\tb}_{[\td}\hat{T}_b^{\ta]}{}_{\tc]}
\end{array}\right)\,,
\eqn{DaTS}
that is, it contains the superfields $P$ as off--diagonal blocks. Moreover, using the algebra of
covariant derivatives, one can easily show at this stage, that the superfields $P$ with upper $SU(8)$
indices are related to the superfields $P$ with lower $SU(8)$ indices by the totally antisymmetric tensor,
\be
P_b{}^{[\ta_1...\ta_4]}\ =\ -\frac{\b}{2}\e^{\ta_1...\ta_8}P_b{}_{[\ta_5...\ta_8]}\,,\qquad\qquad
P_b{}_{[\ta_1...\ta_4]}\ =\ -\frac{\bar{\b}}{2}\e_{\ta_1...\ta_8}P_b{}^{[\ta_5...\ta_8]}\,,
\ee
and the consistency of these two relations fixes the parameters $\b$ and $\bar{\b}$ to be $\b=-\eta/12$
and $\bar{\b}=-\bar{\eta}/12$, with $\eta\bar{\eta}=1$. As a consequence, one has the following
$\eta$--duality relations:
\be
P_b{}^{[\ta_1...\ta_4]}\ =\ \frac{\eta}{4!}\e^{\ta_1...\ta_8}P_b{}_{[\ta_5...\ta_8]}\,,\qquad\qquad
P_b{}_{[\ta_1...\ta_4]}\ =\ \frac{\bar{\eta}}{4!}\e_{\ta_1...\ta_8}P_b{}^{[\ta_5...\ta_8]}\,.
\ee

Concerning the object $\hat{T}_b{}^\tc{}_\ta$, appearing on the diagonal of \equ{DaTS}, is defined as a
function of an a priori arbitrary superfield $T_b{}^\tc{}_\ta$,
\bea
\hat{T}_b{}^\tc{}_\ta &=& T_b{}^\tc{}_\ta+\frac{\vi}{16}\,T_{[\ta\ti\tj]}\s_bT^{[\tc\ti\tj]}\,,
\eea
which in turn appears in the decomposition of the torsion components $T^\tc_\g{}_b{}^\a_\ta$ and
$T_\tc^\dg{}_b{}_\da^\ta$, as it is given by the Bianchi identities $\brg{^\td_\d{}^\dg_\tc{}_b{}^a}$ and
$\brg{_\td^\dd{}_\g^\tc{}_b{}^a}$:
\be
T^\tc_\a{}_b{}^\a_\ta\ =\ -2T_b{}^\tc{}_\ta\,,\qquad\qquad T_\tc^\da{}_b{}_\da^\ta\ =\ 2T_b{}^\ta{}_\tc\,.
\ee
Since this superfield $T_b{}^\tc{}_\ta$ is left undetermined by the constraints we put so far and since
we do not need independent superfields any more in the geometry, we fix it in such a way that we have $0$
on the diagonal of the $(\cd_bT)S$ matrix%
\footnote{Recall that an analysis of conventional constraints on torsion components in extended
supergravity \cite{Mul86b}, \cite{Kis00} shows that the fixing of the traceless part of $T_b{}^\tc{}_\ta$
is nothing but a redefinition of the $SU(8)$ connection component $\F_b{}^\tc{}_\ta$. However, as the
structure group is only $SU(8)$ and not $U(8)$, a fixing of the trace part $T_b{}^\ta{}_\ta$ cannot be
interpreted
as a conventional constraint.}. %
This choice is suggested by the analogy with the matrices $(\cd_\b^\tb T)S$ and $(\cd^\db_\tb T)S$ of
\equ{D1spinTS} and \equ{D2spinTS}. Therefore, we have
\bea
T_b{}^\tc{}_\ta &=& - \frac{\vi}{16}\,T_{[\ta\ti\tj]}\s_bT^{[\tc\ti\tj]}\,,
\eea
and since we also determined the values of the parameters $\a$, $\bar{\a}$, $\b$, $\bar{\b}$ in the
expressions \equ{D1spinTS} and \equ{D2spinTS} of the components with spinorial indices, we can sum up the
results concerning the 1--form $\Omega = (\cd T)S$ as follows:
\bea
\Omega^\td_\d &=&-\frac{1}{2}\frac{1}{3!}\left(
\begin{array}{cc}
0&\eta\e^{\tb\ta\tb'\ta'\td\te\tf\tg}
T_{[\te\tf\tg]\d}\\
\d^{\td\te\tf\tg}_{\tb\ta\tb'\ta'}T_{[\te\tf\tg]\d}&0
\end{array}\right)\,,\label{om1}
\\[2mm]
\Omega_\td^\dd&=&-\frac{1}{2}\frac{1}{3!}\left(
\begin{array}{cc}
0&\d_{\td\te\tf\tg}^{\tb\ta\tb'\ta'}T^{[\te\tf\tg]\dd}\\
\bar{\eta}\e_{\tb\ta\tb'\ta'\td\te\tf\tg}
T^{[\te\tf\tg]\dd}&0
\end{array}\right)\,,\label{om2}
\\[2mm]
\Omega_a &=&-\frac{\vi}{4}\left(
\begin{array}{cc}
0& P_a{}^{[\tb\ta\tb'\ta']}\\
 P_{a[\tb\ta\tb'\ta']}&0
\end{array}\right)\,,
\\[2mm]
\Omega_\au &=&0\,.
\eea
As a consequence, we have to just read out the general matricial structure of the entire form $\OM$:
\be
\OM\ =\ (DT)S\ =\ \left(
\begin{array}{cc}
0&\OM^{[\td\tc\tb\ta]}\\
\OM_{[\td\tc\tb\ta]}&0
\end{array}
\right)
\ee
where the forms $\OM^{[\td\tc\tb\ta]}$ and $\OM_{[\td\tc\tb\ta]}$ are related by the $\eta$--duality
relations
%\footnote{Remark on $\eta$ in \cite{dW79}, page 193: {\it "Notice also that there exist two
%different versions of SO(8) extended supergravity theories, characterized by the duality phase $\eta$
%which can take the values $\pm 1$. In principle these theories are inequivalent, and, for instance, the
%various possible stability groups (i.e. little groups) depend crucially on the duality phase.
%Nevertheless, it is not excluded that the corresponding equations of motion are related. an example of
%such a situation was found in ref..."}}
\be
\OM^{[\th\tg\tf\te]}\ =\ \frac{\eta}{4!}\e^{\th\tg\tf\te\td\tc\tb\ta}\OM_{[\td\tc\tb\ta]}\,,\qquad
\OM_{[\th\tg\tf\te]}\ =\
\frac{\bar{\eta}}{4!}\e_{\th\tg\tf\te\td\tc\tb\ta}\OM^{[\td\tc\tb\ta]}\,,\qquad\eta\bar{\eta}\ =\ 1\,.
\eqn{epsdual}
Let us emphasize here, that although an analogous one--form $\OM$ is present in previous geometrical
descriptions \cite{HL81}, \cite{How82}, its matricial properties like the total antisymmetry of
$\OM_a{}^{[\td\tc\tb\ta]}$ and $\OM_a{}_{[\td\tc\tb\ta]}$ as well as the $\eta$--type duality relations
were imposed \cite{How82} and not deduced from lower dimensional constraints. Concerning our approach we
can say that the total antisymmetry as well as $\eta$--duality of the components of $\OM$, defined as
$\OM\ =\ (DT)S$, have their origin in the requirement that there should be only one type of 1/2 helicity
fields in the multiplet, that is in constraints at 1/2 canonical dimension.

Moreover, as a consequence of its definition, the one--form $\OM$ has further remarkable properties.
First of all, it has the decomposition
\be
\OM\ =\ (DT)S\ =\ (\vd T)S-\F_\su \,,
\eqn{dec}
with $\F_\su$ the $SU(8)$ connection in the $56\times 56$ representation
\be
\F_\su\ =\ \left(
\begin{array}{cc}2\d^{[\td}_{[\tb}\F^{\tc]}{}_{\ta]}&0\\
0&-2\d_{[\td}^{[\tb}\F^{\ta]}{}_{\tc]}
\end{array}\right).
\eqn{fi}
Since it is defined using the covariant derivative, the fact that $\OM$ transforms in a covariant manner
under $SU(8)$ gauge transformations,
\be
\d_X\OM\ =\ -X_\su\OM+\OM X_\su\,,
\ee
with $X_\su$ the gauge parameter in the $56\times 56$ representation \equ{fi}, is obvious. Finally, $\OM$
satisfies the identity
\be
D\OM+\OM\OM+R_\su\ =\ 0\,,
\eqn{DOM}
with again, $R_\su$ the $SU(8)$ curvature in the $56\times 56$ representation \equ{fi}.

All these properties of $\OM$ show that it is a super--analogue of the one--form (usually denoted by $P$)
used to write down the Lagrangian of a non--linear sigma model on $G/K$ coset space with $G$ a
non--compact Lie group and $K$ its maximal compact subgroup. Clearly, the gauge group $SU(8)$ plays the
r\^ole of the compact group $K$, so what remains, is to identify the group $G$. As a matter of fact we
only have access to the properties of a representation of the Lie algebra of $G$, where the object $(\vd
T)S$ takes its values. Observe, that equation \equ{dec} gives the form of the matrix $(\vd T)S$:
\be
(\vd T)S\ =\ \left(
\begin{array}{cc}2\d^{[\td}_{[\tb}\F^{\tc]}{}_{\ta]}&\OM^{[\td\tc\tb\ta]}\\
\OM_{[\td\tc\tb\ta]}&-2\d_{[\td}^{[\tb}\F^{\ta]}{}_{\tc]}
\end{array}\right),\quad\textrm{with}\quad
\OM^{[\th\tg\tf\te]}\ =\ \frac{\eta}{4!}\e^{\th\tg\tf\te\td\tc\tb\ta}\OM_{[\td\tc\tb\ta]}\,,
\ee
where one can recognize the $56\times 56$ representation of the Lie algebra of $E_{7(+7)}$ (see appendix
D of the article \cite{CJ79}). Here the matrix decomposition on the diagonal and off--diagonal parts in
the $28\times 28$ blocks correspond to the decomposition of the Lie algebra of $E_{7(+7)}$ on its Lie
subalgebra $su(8)$ and the orthogonal complement of $su(8)$ with respect to the Killing form. It is
straightforward to verify that the $su(8)$ part corresponds  to the compact, while its complement, the
off--diagonal part, corresponds to the non--compact part of $E_{7(+7)}$. Therefore, we accept that
indeed, the scalars live on the $E_{7(+7)}/SU(8)$ coset space, so we recovered the first silver rule for
the maximally dualised form of the $N=8$ supergravity \cite{Jul98}.

As a final check, one may recall as usual that the dimension of $E_{7(+7)}$ and $SU(8)$ is equal to $133$
and respectively $63$, therefore, the dimension of the coset space parametrized by the scalar fields is
$70$: exactly the number of degrees of freedom in the field strength $\OM_a$ of the scalars, that is
$C_8^4$, and exactly the number of degrees of freedom associated to 0 helicity fields in an $N=8$
supergravity multiplet.

%===================================================================================
\subsection{The $SU(8)$ connection}
%===================================================================================

Recall, that the $SU(8)$ connection 1--form $\F^\tb{}_\ta$ is introduced in the geometry as an a priori
independent object. However, the constraints we imposed imply that it can be expressed as a function of
the scalars and their derivatives. Indeed, the diagonal part of equation \equ{dec} gives immediately the
expression
\be
\F^\tb{}_\ta =\ \frac{1}{3} \left(\vd T^{[\tf\tb]\au} \right)
S_{\au[\tf\ta]} =\ - \frac{1}{3} \left(\vd T_{[\tf\ta]}{}^\au \right) S_\au{}^{[\tf\tb]}\,,
\eqn{fi_su8}
with the property
\be
\left(\vd T^{[\ti\tj]\au} \right) S_{\au[\ti\tj]} =\
\left(\vd T_{[\ti\tj]}{}^\au \right) S_\au{}^{[\ti\tj]} =\ 0\,,
\ee
which ensures that the connection is traceless. Therefore, the $SU(8)$ connection $\F_m{}^\tb{}_\ta$ is
neither an independent field, it can be expressed in terms of the scalar fields of the multiplet and of
their space--time derivatives:
\be
\F^\tb{}_\ta\doubar =\ \vd x^m\F_m{}^\tb{}_\ta =
\frac{1}{3} \vd x^m\left(\partial_m \ct^{[\tf\tb]\au}\right)
\cs_{\au[\tf\ta]} =\ - \frac{1}{3} \vd x^m\left(\partial_m
\ct_{[\tf\ta]}{}^\au \right) \cs_\au{}^{[\tf\tb]}\,,
\eqn{fim_su8}
relation, which is in agreement with the expression given by de Wit and Nicolai %\"{\i}
\cite{dWN82}.

Also, it should be noted that the diagonal part of equation \equ{DOM} expresses the $SU(8)$ curvature in
terms of $\OM$ in a simple way,
\be
R^\tb{}_\ta = - \frac{1}{3} \OM^{[\tb\ti\tj\tk]}
\OM_{[\ta\ti\tj\tk]}\,.
\eqn{R_su8}

The identification of the $SU(8)$ connection was possible because of the vanishing of the central charge
connection \equ{fiuz}. With this condition, the Bianchi identities lead us to the following results: all
the curvature components with at least one lower central charge index as well as the central charge derivatives of
all torsion components vanish,
\be
R_{\az\cc\cb}{}^\ca \ =\ 0 \,, \qquad \cd_\az T_{\cc\cb}{}^\ca\ =\ 0\,.
\ee
Due to the these equations and equation \equ{fiuz}, the central charge sector could appear to be trivial.
Nevertheless, it is essential on the one hand in identifying vectors in the vielbein and the scalars in
the torsion, and on the other hand in deducing the essential properties as the self--duality properties
of the vectors and the non--linear sigma model structure of the scalars.

%%%%%%%%%%%%%%%%%%%%%%%%%%%%%%%%%%%%%%%%%%%%%%%%%%%%%%%%%%%%%%%%%%%%%%%%%%%%%%%%%%%%
\section{Supergravity transformations of the component fields}
%%%%%%%%%%%%%%%%%%%%%%%%%%%%%%%%%%%%%%%%%%%%%%%%%%%%%%%%%%%%%%%%%%%%%%%%%%%%%%%%%%%%

Once the component fields of the supergravity multiplet are identified, the aim of the present section is
to deduce their supergravity transformations in terms of component fields and compare these
transformations with those found on the component level \cite{CJ79}, \cite{dWN82}.

Recall that one of the fundamental advantages of the geometrical description in central charge superspace
is that space-time diffeomorphisms, supersymmetry and gauge transformations identified as central charge
transformations are treated on the same footing as superspace diffeomorphisms. Given that component fields
were identified in the super vielbein and in torsion components and also that we know how such geometrical
objects transform under supergravity transformations \equ{dframe}, it is a straightforward exercice to
write down how these transformations act on components.

Let us begin with the component fields identified in the frame \equ{frame}: the graviton, the gravitini
and graviphotons. Their supergravity transformations can be read out of the explicit component expansion
\be
\dwz_\x E_m{}^\ca\loco\ =\ \cd_m\x^\ca\loco+E_m{}^\cb\x^\cc T_{\cc\cb}{}^\ca\loco
\eqn{dEma}
of the transformation \equ{dframe} taking the lowest superfield component (noted here by the bar $\loco$)
and choosing for $\ca$ either vector, spinorial or respectively central charge indices. As for the scalars
and spinor fields, identified in torsion components, that is as the lowest superfield component of
super--covariant fields $V$, their supergravity transformation is simply
\be
\dwz_\x V\loco\ =\ \cl_\x V\loco\ =\ \x^\ca\cd_\ca V\loco\,.
\eqn{dV}

However, in order to make explicit these transformation laws, we need the expression in component fields of
the lowest superfield components of the basic superfields appearing in torsion components of \equ{dEma}
and covariant derivatives of \equ{dV}. The aim of the next subsection is precisely to give the list of the
necessary expansions.

%%%%%%%%%%%%%%%%%%%%%%%%%%%%%%%%%%%%%%%%%%%%%%%%%%%%%%%%%%%%%%%%%
\subsection{Supercovariant$\rightarrow$component toolkit}
%%%%%%%%%%%%%%%%%%%%%%%%%%%%%%%%%%%%%%%%%%%%%%%%%%%%%%%%%%%%%%%%%
\label{sup_comp}

The basic superfields appearing in torsion components (see appendix \ref{sol}) are all super--covariant
quantities and their lowest superfield components inherit this property. Let us sum up in this paragraphe
the component expressions of the supercovariant field strengths, needed to write down supergravity
transformations of the component fields. General formulas used to determinate these expressions are
easily written using the notation $E^\ca\doubar=e^\ca=\vd x^m e_m{}^\ca$ \cite{BGG01}.

Recall that the graviton, gravitini and graviphotons are identified in the super-vielbein. Thus, their
field strength can be found in their covariant counterparts using
\be
T^\ca\doubar\  =\  \frac{1}{2}\vd x^m \vd x^n\left(\cd_ne_m{}^\ca-\cd_me_n{}^\ca\right)
\  =\  \frac{1}{2}e^\cb e^\cc T_{\cc\cb}{}^\ca\loco\,.
\ee
For $\ca=a$ one finds the relation
\be
\cd_ne_m{}^a-\cd_m e_n{}^a\ =\ i\p_{[n\ta}\s^a\bar{\p}_{m]}{}^\ta\,,
\eqn{Tcba}
which determinates the Lorentz connection $\F_m{}_{kl} = e_k{}^be_l{}^a\F_m{}_{ba}$ in terms of the
vierbein, its derivatives and
gravitini fields%
\footnote{One may observe that in \cite{dWN82}, page 334, the Lorentz connection depends also on the
spinor fields. However, this difference is just a matter of redefinition, it corresponds in our
geometrical description to the replacement of the conventional constraint $T_{cb}{}^a=0$ by
$T_{cb}{}^a=\la\e_{dcb}{}^a(T_{[\tc\tb\ta]}\s^dT^{[\tc\tb\ta]})$.}%
\bea
\F_m{}_{kl}& =&
\frac{1}{2}\left(e_m{}^a\partial_ke_{la}-e_l{}^a\partial_me_{ka}-e_k{}^a\partial_le_{ma}\right)
-(k\leftrightarrow l)\nn\\[2mm]
&&-\frac{\vi}{4}\left(\p_m{}_\ta\s_k\bp_l{}^\ta-\p_l{}_\ta\s_m\bp_k{}^\ta+\p_k{}_\ta\s_l\bp_m{}^\ta\right)
-(k\leftrightarrow l)\,.
\label{fimba}
\eea

For $\ca=\au$, the central charge indices, we obtain in general the covariant field strength of the
graviphotons
\bea
F_{ba}{}^\au\loco &=& e_b{}^n e_a{}^m\cf_{nm}{}^\au
+e_b{}^n e_a{}^m\left[\frac{1}{4}(\bar{\p}_{n}{}^{\tc}\bar{\p}_{m}{}^{\tb})
+\frac{\vi}{8}(\p_{[n}{}_{\ta}\s_{m]}\T^{[\tc\tb\ta]})\right]\ct_{[\tc\tb]}{}^{\au}\nn\\[2mm]
&&
+e_b{}^n e_a{}^m\left[\frac{1}{4}(\p_{n}{}_{\tc}\p_{m}{}_{\tb})
+\frac{\vi}{8}(\bp_{[n}{}^{\ta}\bs_{m]}\T_{[\tc\tb\ta]})\right]\ct^{[\tc\tb]}{}^{\au}\,,
\eea
with $\cf_{nm}{}^\au$ the field strength of the graviphotons $\cf_{nm}{}^\au=\partial_{n}v_{m}{}^\au
-\partial_{m}v_{n}{}^\au$. This can also be written in the $SU(8)$ basis:
\bea
F_{ba}{}_{[\tc\tb]}\loco&=&e_b{}^n e_a{}^m\cf_{nm}{}^\au \cs_\au{}_{[\tc\tb]}
+e_{[b}{}^ne_{a]}{}^m\left[\frac{1}{4}(\p_{n}{}_{\tc}\p_{m}{}_{\tb})
+\frac{\vi}{8}(\bp_{n}{}^{\ta}\bs_{m}\T_{[\tc\tb\ta]})\right]
\,,\\[2mm]
F_{ba}{}^{[\tc\tb]}\loco&=&e_b{}^n e_a{}^m\cf_{nm}{}^\au \cs_\au{}^{[\tc\tb]}
+e_{[b}{}^n e_{a]}{}^m\left[\frac{1}{4}(\bar{\p}_{n}{}^{\tc}\bar{\p}_{m}{}^{\tb})
+\frac{\vi}{8}(\p_{n}{}_{\ta}\s_{m}\T^{[\tc\tb\ta]})\right]\,.
\eea
However, as we already have seen in section \ref{selfdphoton}, the field strength of the graviphotons
satisfy self--duality relations, which reduce their degrees of freedom to the half. We have seen in
particular, that the dynamic part of this field strength corresponds to the self--dual part
$\fp_{ba}{}_{[\tb\ta]}$ of the component $F_{ba}{}_{[\tb\ta]}$ and the anti--self--dual part
$\fm_{ba}{}^{[\tb\ta]}$ of the component $F_{ba}{}^{[\tb\ta]}$, while the remaining parts (the
anti--self--dual part of the component $F_{ba}{}_{[\tb\ta]}$ and the self--dual part of the component
$F_{ba}{}^{[\tb\ta]}$) are given as quadratic terms in the spinor superfields \equ{f1},
\equ{f2}. Therefore, we are rather interested in the expression of lowest superfield components of the
objects $\fp_{ba}{}_{[\tb\ta]}$ and $\fm_{ba}{}^{[\tb\ta]}$:
\bea
\fp_{ba}{}_{[\tc\tb]}\loco&=&e_b{}^n e_a{}^m\cf^{\textrm{{\tiny{+}}}}_{nm}{}^\au \cs_\au{}_{[\tc\tb]}
-\frac{1}{8}\textrm{tr}(\s_{ba}\s^{nm})\left[\left(\p_{n}{}_{\tc}\p_{m}{}_{\tb}\right)
+\frac{\vi}{2}\left(\bp_{n}{}^{\ta}\bs_{m}\T_{[\tc\tb\ta]}\right)\right],\\[2mm]
\fm_{ba}{}^{[\tc\tb]}\loco&=&e_b{}^n e_a{}^m\cf^-_{nm}{}^\au \cs_\au{}^{[\tc\tb]}
-\frac{1}{8}\textrm{tr}(\bs_{ba}\bs^{nm})\left[\left(\bar{\p}_{n}{}^{\tc}\bar{\p}_{m}{}^{\tb}\right)
+\frac{\vi}{2}\left(\p_{n}{}_{\ta}\s_{m}\T^{[\tc\tb\ta]}\right)\right].
\eea
These are the objects, which correspond to the super--covariant field strength of the gravi\-photons
$\hfp_{nm}{}_{[\tc\tb]}$ and $\hfm_{nm}{}^{[\tc\tb]}$ used on the component level in the article
\cite{dWN82}, which can be defined as
\be
\hfp_{nm}{}_{[\tc\tb]}\ =\ e_n{}^be_m{}^a\fp_{ba}{}_{[\tc\tb]}\loco\,,\qquad
\hfm_{nm}{}^{[\tc\tb]}\ =\ e_n{}^be_m{}^a\fm_{ba}{}^{[\tc\tb]}\loco\,,
\ee
and have the expressions
\bea
\hfp_{nm}{}_{[\tc\tb]}&=&\cf^{\textrm{{\tiny{+}}}}_{nm}{}^\au \cs_\au{}_{[\tc\tb]}
-\frac{1}{8}\textrm{tr}(\s_{nm}\s^{kl})\left[\left(\p_{k}{}_{\tc}\p_{l}{}_{\tb}\right)
+\frac{\vi}{2}\left(\bp_{k}{}^{\ta}\bs_{l}\T_{[\tc\tb\ta]}\right)\right],\\[2mm]
\hfm_{nm}{}^{[\tc\tb]}&=&\cf^-_{nm}{}^\au \cs_\au{}^{[\tc\tb]}
-\frac{1}{8}\textrm{tr}(\bs_{nm}\bs^{kl})\left[\left(\bar{\p}_{k}{}^{\tc}\bar{\p}_{l}{}^{\tb}\right)
+\frac{\vi}{2}\left(\p_{k}{}_{\ta}\s_{l}\T^{[\tc\tb\ta]}\right)\right].
\eea
In order to be able to compare our results to those of the component approach in \cite{dWN82}, we will
systematically use the objects $\hfp_{nm}{}_{[\tc\tb]}$ and $\hfm_{nm}{}^{[\tc\tb]}$ in our component
formulas.

As for $\ca=_\a^\ta$ and $\ca=^\da_\ta$ we have the expression of the covariant field strength of the
gravitini
\bea
T_{cb}{}_\ta^\a\loco &=& e_b{}^me_c{}^n\cd_{[n}\p_{m]}{}^\a_\ta
-4\vi e_{[b}{}^n F_{c]a}{}_{[\tb\ta]}\loco(\bp_n{}^\tb\bs^a)^\a
+\frac{1}{4}e_b{}^me_c{}^n(\bar{\p}_n{}^{\tc}\bar{\p}_m{}^{\tb})\T_{[\tc\tb\ta]}{}^\a\nn\\[2mm]
&&
-\frac{\vi}{16}e_{[b}{}^n(\p_n{}_\tb\s_{c]}\bs_a)^\a (\T_{[\ta\te\tf]}\s^a\T^{[\tb\te\tf]})
+\frac{\vi}{48}e_{[b}{}^n(\p_n{}_\ta\s_{c]a})^\a (\T_{[\te\tf\tg]}\s^a\T^{[\te\tf\tg]})\,,\nn
\eea
\bea
T_{cb}{}^\ta_\da\loco &=& e_b{}^me_c{}^n\cd_{[n}\bp_{m]}{}_\da^\ta
-4\vi e_{[b}{}^n F_{c]a}{}^{[\tb\ta]}\loco(\p_n{}_\tb\s^a)_\da
+\frac{1}{4}e_b{}^me_c{}^n({\p}_n{}_{\tc}{\p}_m{}_{\tb})\T^{[\tc\tb\ta]}{}_\da\nn\\[2mm]
&&
+\frac{\vi}{16}e_{[b}{}^n(\bp_n{}^\tb\bs_{c]}\s_a)_\da (\T_{[\tb\te\tf]}\s^a\T^{[\ta\te\tf]})
-\frac{\vi}{48}e_{[b}{}^n(\bp_n{}^\ta\bs_{c]a})_\da (\T_{[\te\tf\tg]}\s^a\T^{[\te\tf\tg]})\,.\nn
\eea
Defining $\hat{\Psi}_{nm}{}^\a_\ta=e_n{}^ce_m{}^b T_{cb}{}_\ta^\a\loco$ and respectively
$\hat{\Psi}_{nm}{}_\da^\ta=e_n{}^ce_m{}^b T_{cb}{}^\ta_\da\loco$, one obtains the supercovariant field
strength of the gravitini used in the component formalism in the article \cite{dWN82}:
\bea
\hat{\Psi}_{nm}{}^\a_\ta&=&\cd_{[n}\p_{m]}{}^\a_\ta
+\frac{1}{4}(\bar{\p}_n{}^{\tc}\bar{\p}_m{}^{\tb})\T_{[\tc\tb\ta]}{}^\a\nn\\[2mm]
&&+4\vi (\bp_{[n}{}^\tb\bs^l)^\a \left[\hat{F}^{+}_{m]l}{}_{[\tb\ta]}
-\frac{\eta}{2(4!)^2}\e_{\tb\ta\tf_1...\tf_6}(\T^{[\tf_1\tf_2\tf_3]}\bs_{m]l}\T^{[\tf_4\tf_5\tf_6]})\right]\nn\\[2mm]
&&
+\frac{\vi}{16}(\p_{[n}{}_\tb\s_{m]}\bs_l)^\a (\T_{[\ta\te\tf]}\s^l\T^{[\tb\te\tf]})
-\frac{\vi}{48}(\p_{[n}{}_\ta\s_{m]l})^\a (\T_{[\te\tf\tg]}\s^l\T^{[\te\tf\tg]})\,,\nn
\eea
\bea
\hat{\Psi}_{nm}{}_\da^\ta&=&\cd_{[n}\bp_{m]}{}_\da^\ta
+\frac{1}{4}(\p_n{}_{\tc}\p_m{}_{\tb})\T^{[\tc\tb\ta]}{}_\da\nn\\[2mm]
&&+4\vi (\p_{[n}{}_\tb\s^l)_\da \left[\hat{F}^{-}_{m]l}{}^{[\tb\ta]}
-\frac{\bar{\eta}}{2(4!)^2}\e^{\tb\ta\tf_1...\tf_6}(\T_{[\tf_1\tf_2\tf_3]}\s_{m]l}\T_{[\tf_4\tf_5\tf_6]})\right]\nn\\[2mm]
&&
-\frac{\vi}{16}(\bp_{[n}{}^\tb\bs_{m]}\s_l)_\da (\T_{[\tb\te\tf]}\s^l\T^{[\ta\te\tf]})
+\frac{\vi}{48}(\bp_{[n}{}^\ta\bs_{m]l})_\da (\T_{[\te\tf\tg]}\s^l\T^{[\te\tf\tg]})\,.\nn
\eea

Finally, for the field strength of the scalars $\OM_a$, we can use the definition of the double
projection on the one--form $\OM$,
\be
\OM\doubar\ =\ \vd x^m\OM_m \ =\ e^\ca\OM_\ca\loco\,,
\ee
and obtain
\be
\OM_a\loco\ =\ e_a{}^m\left[\OM_m
-\frac{1}{2}\p_m{}^\a_\ta\OM^\ta_\a\loco
-\frac{1}{2}\bp_m{}_\da^\ta\OM_\ta^\da\loco\right]\,,
\ee
with $\OM_m\ =\ (\cd_m\ct)\cs$ the ordinary field strength of the scalars. Then, using the expressions of the
matrix components in \equ{om1}, \equ{om2}, we obtain for the two off--diagonal blocks
\bea
\OM_a{}^{[\td\tc\tb\ta]}\loco&=&e_a{}^m\left[\OM_m{}^{[\td\tc\tb\ta]}
+\frac{\eta}{4!}\e^{\td\tc\tb\ta\te\tf\tg\th}\left(\p_m{}_\te \T_{[\tf\tg\th]}\right)
+\left(\bp_m{}^{[\td} \T^{[\tc\tb\ta]]}\right)\right],
\\[2mm]
\OM_a{}_{[\td\tc\tb\ta]}\loco&=&e_a{}^m\left[\OM_m{}_{[\td\tc\tb\ta]}
+\left(\p_m{}_{[\td} \T_{[\tc\tb\ta]]}\right)
+\frac{\bar{\eta}}{4!}\e_{\td\tc\tb\ta\te\tf\tg\th}\left(\bp_m{}^\te \T^{[\tf\tg\th]}\right)\right].
\eea
Analogously to the previous case, the super--covariant field strength of the scalars is defined as
$\hOM_m=e_m{}^a\OM_a\loco$ and has the component expansions
\bea
\hOM_m{}^{[\td\tc\tb\ta]}&=&\OM_m{}^{[\td\tc\tb\ta]}
+\frac{\eta}{4!}\e^{\td\tc\tb\ta\te\tf\tg\th}\left(\p_m{}_\te \T_{[\tf\tg\th]}\right)
+\left(\bp_m{}^{[\td} \T^{[\tc\tb\ta]]}\right),
\\[2mm]
\hOM_m{}_{[\td\tc\tb\ta]}&=&\OM_m{}_{[\td\tc\tb\ta]}
+\left(\p_m{}_{[\td} \T_{[\tc\tb\ta]]}\right)
+\frac{\bar{\eta}}{4!}\e_{\td\tc\tb\ta\te\tf\tg\th}\left(\bp_m{}^\te \T^{[\tf\tg\th]}\right).
\eea

We are now ready to explicite supergravity transformations  of the component fields.

%===================================================================================
\subsection{Supersymmetry transformations}
%===================================================================================

Recall that in the superspace description supersymmetry transformations are supergravity transformations
with only spinorial non--zero parameters
\be
\x^\ca = (0,\x^\a_\ta,\bx^\ta_\da,0)\,.
\ee
 Therefore, using the
general expressions \equ{dEma}, \equ{dV} as well as the expressions of torsion components and spinorial
derivatives of the basic fields summed up in appendix \ref{sol}, we have the following component
transformation laws.

For the graviton, gravitini and graviphotons we have
\be
\dwz_{\x}e_m{}^a\ =\ \vi\left(\x_\ta\s^a\bp_m{}^\ta+\bx^\ta\bs^a\p_m{}_\ta\right),
\ee
\bea
\frac{1}{2}\dwz_{\x}\p_m{}^\a_\ta&=&\cd_m \x^\a_\ta
-\frac{1}{2}(\bx^\tc\bp_m{}^\tb) \T_{[\tc\tb\ta]}{}^\a\nn\\[2mm]
&&
+4\vi(\bx^\tc\bs^n)^\a \hfp_{nm}{}_{[\tc\ta]}
+\frac{2\vi\eta}{(4!)^2}\e_{\ta\tc\tf_1...\tf_6}
(\bx^\tc \T^{[\tf_1\tf_2\tf_3]})(\T^{[\tf_4\tf_5\tf_6]}\bs_m)^\a\nn\\[2mm]
&&
-\frac{\vi}{16}(\x_\tf\s_m\bs_n)^\a \ (\T_{[\ta\tc\tb]}\s^n\T^{[\tf\tc\tb]})
+\frac{\vi}{48}(\x_\ta\s_{mn})^\a \ (\T_{[\tf\tc\tb]}\s^n\T^{[\tf\tc\tb]})\,,
\\[2mm]
\frac{1}{2}\dwz_{\x}\bp_m{}_\da^\ta&=&\cd_m \bx_\da^\ta
-\frac{1}{2}(\x_\tc\p_m{}_\tb) \T^{[\tc\tb\ta]}{}_\da\nn\\[2mm]
&&
+4\vi(\x_\tc\s^n)_\da \hfm_{nm}{}^{[\tc\ta]}
+\frac{2\vi\bar{\eta}}{(4!)^2}\e^{\ta\tc\tf_1...\tf_6}
(\x_\tc \T_{[\tf_1\tf_2\tf_3]})(\T_{[\tf_4\tf_5\tf_6]}\s_m)_\da\nn\\[2mm]
&&
+\frac{\vi}{16}(\bx^\tf\bs_m\s_n)_\da \ (\T_{[\tf\tc\tb]}\s^n\T^{[\ta\tc\tb]})
-\frac{\vi}{48}(\bx^\ta\bs_{mn})_\da \ (\T_{[\tf\tc\tb]}\s^n\T^{[\tf\tc\tb]})\,,
\eea
\bea
\dwz_\x v_m{}^\au&=&-\frac{1}{2}\left[(\x_\tb\p_m{}_\ta)
+\frac{\vi}{4}(\bx^\tc\bs_m \la_{[\tc\tb\ta]})\right]\ct^{[\tb\ta]}{}^\au\nn\\[2mm]
&&
-\frac{1}{2}\left[(\bx^\tb\bp_m{}^\ta)
+\frac{\vi}{4}(\x_\tc\s_m \la^{[\tc\tb\ta]})\right]\ct_{[\tb\ta]}{}^\au\,.\label{susy_vect}
\eea

Using the spinorial derivatives \equ{derspin1}--\equ{derspin2} of the gravigini superfields we obtain for
the helicity 1/2 fields the transformation law
\bea
\dwz_{\x} \T_{[\tc\tb\ta]}{}^\a&=&
4\d^{\td\te\tf}_{\tc\tb\ta}(\x_\td\s^{mn})^\a \hfp_{mn}{}_{[\te\tf]}
+4\vi(\bx^\td\bs^m)^\a\hOM_m{}_{[\td\tc\tb\ta]}\nn\\[2mm]
&&-\frac{\bar{\eta}}{4!}\e_{\tc\tb\ta\te_1...\te_5}
\x_\td^\a\ (\T^{[\td\te_1\te_2]}\T^{[\te_3\te_4\te_5]})\,,\\[2mm]
\dwz_{\x} \T^{[\tc\tb\ta]}{}_\da&=&
4\d_{\td\te\tf}^{\tc\tb\ta}(\bx^\td\bs^{mn})_\da \hfm_{mn}{}^{[\te\tf]}
+4\vi(\x_\td\s^m)_\da\hOM_m{}^{[\td\tc\tb\ta]}\nn\\[2mm]
&&-\frac{{\eta}}{4!}\e^{\tc\tb\ta\te_1...\te_5}
\bx^\td_\da\ (\T_{[\td\te_1\te_2]}\T_{[\te_3\te_4\te_5]})\,.
\eea

As for the supersymmetry transformations of the scalar fields, one obtains
\be
\dwz_\x T\loco\ =\ \cl_\x T\loco\ =\ (\imath_\x\OM)\,T\loco\,,
\ee
that is, just a rotation by a matrix $\S$, which is an element of the orthogonal complement of the Lie
algebra $su(8)$ in the Lie algebra of $E_{7(+7)}$,
\be
\dwz_\x \ct\ =\ \S\,\ct\,,\quad\textrm{with}\quad
\S\ =\ \imath_\x\OM\loco\ =\ \left(\begin{array}{cc}
0&\S^{[\td\tc\tb\ta]}\\
\S_{[\td\tc\tb\ta]}&0
\end{array}\right),
\eqn{dscalaire}
or in matrix components,
\be
\dwz_\x \ct^{[\td\tc]}{}^\au\ =\ \S^{[\td\tc\tb\ta]}\ct_{[\tb\ta]}{}^\au\,,\qquad
\dwz_\x \ct_{[\td\tc]}{}^\au\ =\ \S_{[\td\tc\tb\ta]}\ct^{[\tb\ta]}{}^\au\,.
\ee
Here, of course, the objects $\S^{[\td\tc\tb\ta]}$ and $\S_{[\td\tc\tb\ta]}$ are related by the duality
relation
\equ{epsdual} and they are computed using the component expressions \equ{om1} and
\equ{om2} of $\OM_\a^\ta$ and $\OM^\da_\ta$:
\bea
\S^{[\td\tc\tb\ta]}&=&-2\left[\frac{\eta}{4!}\e^{\td\tc\tb\ta\te\tf\tg\th}\,\left(\x{}_\te \T_{[\tf\tg\th]}\right)
+\left(\bx{}^{[\td} \T^{[\tc\tb\ta]]}\right)\right],\\[2mm]
\S_{[\td\tc\tb\ta]}&=&
-2\left[\left(\x_{[\td} \T_{[\tc\tb\ta]]}\right)
+\frac{\bar{\eta}}{4!}\e_{\td\tc\tb\ta\te\tf\tg\th}\,\left(\bx^\te \T^{[\tf\tg\th]}\right)\right].
\eea

Finally, we can also notice, that as pointed out on the component level in the article \cite{dWN82}, the
supersymmetry transformations of the $SU(8)$ connection, which is a function of the scalar fields and
their derivatives \equ{fi_su8}, can also be given in a simple way using the above defined
field--dependent parameters $\S$. Indeed, using the expression \equ{R_su8} of the $SU(8)$ curvature, we
have
\be
\dwz_\x\F^\tb{}_\ta\ =\ \imath_\x R^\tb{}_\ta\ =\
\frac{1}{3}\left[\S^{[\tb\te\tf\tg]}\OM_{[\ta\te\tf\tg]}
-\S_{[\ta\te\tf\tg]}\OM^{[\tb\te\tf\tg]}\right],
\ee
while on the component level,
\be
\dwz_\x\F_m{}^\tb{}_\ta\ =\
\frac{1}{3}\left[\S^{[\tb\te\tf\tg]}\OM_m{}_{[\ta\te\tf\tg]}
-\S_{[\ta\te\tf\tg]}\OM_m{}^{[\tb\te\tf\tg]}\right].
\ee

These transformation laws are in perfect concordance with those found at the component 
level in \cite{dWN82} (we prefer to make reference to this work, because the
transformation laws given there contain all the non--linear terms even in the spinor
fields, which are not given explicitly in the original works like \cite{CJ79}).

%===================================================================================
\subsection{Central charge transformations}
%===================================================================================

%{\it Question: comment expliquer que nous avons a priori 56 parametres de jauge $\z^\au$ (ou dans l'autre
%base, c'est pareil), mais on a 28 vecteurs. Quest-ce que ca veut dire: tf. de jauge avec un champ de jauge
%self-dual?}

In the geometrical description central charge transformations are just supergravity transformations in
the direction of central charge coordinates, that is, with parameters
\be
\z^\ca=(0,0,0,\z^\au)\,.
\ee
Therefore, we can use the general formulas \equ{dEma} and \equ{dV} with these parameters in order to give
these transformations. However, as a consequence of the constraint $T_{\az\cb}{}^\ca=0$ as well as of the
fact that the central charge derivative of all torsion components vanish, $\cd_\az T_{\cc\cb}{}^\ca=0$,
all component fields but the graviphotons transform trivially under central charge transformations.

The transformation of the graviphotons is simply
\be
\dwz_{\z} v_m{}^\au\ =\ \cd_m \z^\au\,,
\eqn{cc_vect}
since they are the gauge fields corresponding the central charge transformations.

Also, it was noticed in \cite{dWN82}, that a gauge transformation with a scalar--dependent parameter
appears in the commutator of two supersymmetry transformations. This feature appears naturally in our
approach, since one handles with central charge transformations which are present in the algebra of
supergravity transformations with a parameter which depends on the 0 dimensional torsion components,
where the scalars were identified. In order to see this in detail, recall that the commutator of two
supergravity transformations acting on the frame is
\be
\left[\dwz_{\x},\dwz_{\eta}\right]E^\ca
\ =\ \dwz_{[\x,\eta]}E^\ca-E^\cb \imath_{\x}\imath_{\eta} R_\cb{}^\ca\,,
\ee
with
\be
\left[\x,\eta\right]^\ca\ =\ \x^\cb\eta^\cc T_{\cc\cb}{}^\ca
+\x^\cb(\cd_\cb \eta^\ca)-\eta^\cb(\cd_\cb\x^\ca).
\ee
Then choosing for instance the only non--zero parameters $\x^\a_\ta\loco$ and $\eta^\a_\ta\loco$, the
commutator on a graviphoton becomes
\bea
\left[\dwz_{\x},\dwz_{\eta}\right]v_m{}^\au
& =&\cd_m\left(\ (\x_\tb\eta_\ta)\  \ct^{[\tb\ta]\au}\ \right)
-\frac{\vi}{8}(\x_\tb\eta_\ta)\ (\T^{[\tb\ta\tf]}\bs_m\T_{[\tf\td\tc]})\ct^{[\td\tc]\au}\nn\\[2mm]
&&
-\frac{1}{2}(\x_\tb\eta_\ta)\ (\T^{[\tb\ta\td]}\bp_m{}^\tc)  \ct_{[\td\tc]}{}^\au\,,
\eea
and indeed, it contains a central charge transformation \equ{cc_vect} with parameter $(\x_\tb\eta_\ta)
\ct^{[\tb\ta]\au}$ depending on scalars as well as a supersymmetry transformation \equ{susy_vect} with parameter
$(\x_\tb\eta_\ta) \T^{[\tb\ta\tf]}{}_\da$ depending on the helicity 1/2 fields.

%%%%%%%%%%%%%%%%%%%%%%%%%%%%%%%%%%%%%%%%%%%%%%%%%%%%%%%%%%%%%%%%
\section{The equations of motion}
%%%%%%%%%%%%%%%%%%%%%%%%%%%%%%%%%%%%%%%%%%%%%%%%%%%%%%%%%%%%%%%%

The problem of the derivation of field equations of motion without the knowledge of a Lagrangian, using
considerations on representations of the symmetry group, was considered for a long time \cite{Bel74},
\cite{Wei95}. The question is particularly interesting for supersymmetric theories and for this case
various approaches have been developed. Here we use the techniques of superspace geometry introduced by
Wess and Zumino, which consist in looking to consequences of covariant constraints corresponding to
on--shell field content of a representation of the supersymmetry algebra.

Indeed, the next and last step in the geometric description of the $N=8$ on--shell supergravity theory is
to deduce the equations of motion implied by the constraints we used to identify the multiplet in the
geometry, and compare them with those found from the Lagrangian given in the original works in the
component formalism \cite{CJ79}, \cite{dWN82}.

The method of deducing the equations of motion for $N\geq 3$ extended supergravity is similar to the case
of the $N=1$ Yang--Mills theory in the sense that the gravigino superfields $T_{[\tc\tb\ta]\a}$,
$T^{[\tc\tb\ta]\da}$ play an analogous r\^ole to the gaugino superfields and all equations of motion but
those for the graviton and gravitini are contained in their higher superfield components \cite{GK02}.
Therefore, these equations of motion are found by successively acting with spinorial derivatives on the
spinorial derivatives of the gravigini superfields. This is the approach which was adopted in previous
superspace descriptions of the $N=8$ supergravity in order to derive the {\it free} equations of motion
from the geometry \cite{Sie81b}, \cite{HL81} (see also references \cite{GG83} and \cite{CL86}).
Alternatively, one could also just pick out certain Bianchi identities, which give the equations of
motion for the component fields. This strategy to obtain equations of motion is outlined in the article
\cite{BH79}. For the purpose of putting in evidence free equations of motion of component fields it is
sufficient to consider only the linearized version and the calculations are simple. However, one has to
consider the full theory if one wants to obtain all the {\it non--linear terms} which arise in equations
of motion derived from a Lagrangian in component formalism.

%%===================================================================
%\subsection{Equations of motion in terms of basic superfields}
%%===================================================================

To begin with let us explain in detail how one can derive the equations of motion for the gravigini.
First, recall that the spinorial derivatives of the gravigino superfields $T_{[\tc\tb\ta]\a}$ (see
appendix \ref{sol}) have for instance the properties
\bea
\cd^{\td\a}T_{[\tc\tb\ta]\a}&=&\frac{\bar{\eta}}{12}\e_{\tc\tb\ta\ti\tj\tk\tl\tm}
\left( T^{[\td\ti\tj]}T^{[\tk\tl\tm]}\right)\,,\label{dspin1}\\[2mm]
\sum_{\td\tc}\cd_\td^\dd T_{[\tc\tb\ta]\a}& =& 0\,.\label{dspin2}
\eea
In order to get the Dirac equation of the 1/2 helicity field, one can just act on the last relation by
the spinorial derivative $\cd^\te_\e$ obtaining
\be
\sum_{\td\tc}\left(\left\{\cd^\te_\e,\cd_\td^\dd\right\} T_{[\tc\tb\ta]\a}
-\cd_\td^\dd\left(\cd^\te_\e T_{[\tc\tb\ta]\a}\right)\right)\ =\ 0,
\ee
and take the antisymmetric part of this relation in the indices $\e$ and $\a$. Then, using the algebra of
covariant derivatives as well as equation \equ{dspin1} and again the expressions of the spinorial
derivatives of the gravigini superfields one obtains
\bea
\cd^{\a\da}T_{[\tc\tb\ta]\a}&=&
\frac{\vi\bar{\eta}}{3}\e_{\tc\tb\ta\te\tf\tg\th\ti}
\fm_{ba}{}^{[\te\tf]}(\bs^{ba}T^{[\tg\th\ti]})^\da\nn\\
[2mm]
&&+\frac{\vi}{16}(T_{[\tc\tb\ta]}T_{[\te\tf\tg]})T^{[\te\tf\tg]\da}
-\frac{\vi}{8}\oint_{\tc\tb\ta}(T_{[\tc\te\tf]}T_{[\tb\ta\tg]})T^{[\te\tf\tg]\da}\,.
\label{em_spin_u}
\eea
Alternatively, it is worthwhile to observe that this equation of motion can also be deduced from the
Bianchi identity $\brg{^\dd_\td{}_c{}^\db_\tb{}^\a_\ta}$ of dimension 3/2 using the expressions of
torsion components and of spinorial derivatives (see appendix \ref{sol}) of lower canonical dimension.

The conjugate Dirac equation can be obtained in an analogous way:
\bea
\cd_{\a\da}T^{[\tc\tb\ta]\da}&=&
\frac{\vi\eta}{3}\e^{\tc\tb\ta\te\tf\tg\th\ti}
\fp_{ba}{}_{[\te\tf]}(\s^{ba}T_{[\tg\th\ti]})_\a\nn\\
[2mm]
&&+\frac{\vi}{16}(T^{[\tc\tb\ta]}T^{[\te\tf\tg]})T_{[\te\tf\tg]\a}
-\frac{\vi}{8}\oint^{\tc\tb\ta}(T^{[\tc\te\tf]}T^{[\tb\ta\tg]})T_{[\te\tf\tg]\a}\,.
\label{em_spin_d}
\eea

Moreover, the equations of motion for the graviphotons and those of the scalars can be deduced by further
acting with covariant spinorial derivatives on the equations of motion \equ{em_spin_u} and
\equ{em_spin_d} of the gravigini fields as follows.

On the one hand the trace part in the $SU(8)$ indices of the spinorial derivative $\cd^\td_\d$ of the
Dirac equation \equ{em_spin_u} gives the equations of motion for the self--dual field strength
$\fp_{ba}{}_{[\tb\ta]}$ of the graviphotons,
\bea
\cd_f \fp{}^{fa}{}_{[\tb\ta]} &=& -\cd_f\left[\frac{\bar{\eta}}{2(4!)^2} \e_{\tb\ta\tf_1...\tf_6}
 \left(T^{[\tf_1\tf_2\tf_3]} \bs^{fa} T^{[\tf_4\tf_5\tf_6]} \right)\right]
\nn\\
& &+\OM_f{}_{[\tb\ta\ti\tj]}\left[ \fm{}^{fa}{}^{[\ti\tj]} + \frac{\eta}{2(4!)^2}
\e^{\ti\tj\tf_1...\tf_6} \left(T_{[\tf_1\tf_2\tf_3]}\s^{fa}T_{[\tf_4\tf_5\tf_6]} \right) \right]
\nn\\
 & &-\frac{1}{4(4!)} \left(T_{[\tf\tb\ta]} \s^f \bs^a T_{[\ti\tj\tk]} \right) \OM_f{}^{[\tf\ti\tj\tk]}
+ \frac{\vi}{16} \left(T_{[\tf\tb\ta]} \s^{fg} \s^a T^{[\tf\ti\tj]} \right) \fp_{fg}{}_{[\ti\tj]}
\nn\\
 & &+\frac{\vi \bar{\eta}}{2(4!)^3} \e_{\ti\tj\tk\tl\tm\te\tf\tg}
\left( T^{[\td\tl\tm]}T^{[\te\tf\tg]} \right) \left(T_{[\td\tb\ta]}
\s^a T^{[\ti\tj\tk]} \right) \,,
\label{em_vect_d}
\eea
while the conjugate field equations for $\fm_{ba}{}_{[\tb\ta]}$ can be obtained in an analogous way,
\bea
\cd_f \fm{}^{fa}{}^{[\tb\ta]} &=&- \cd_f \left[\frac{\eta}{2(4!)^2} \e^{\tb\ta\tf_1...\tf_6}
\left(T_{[\tf_1\tf_2\tf_3]} \s^{fa} T_{[\tf_4\tf_5\tf_6]} \right)\right]
\nn\\
& & +\OM_f{}^{[\tb\ta\ti\tj]}\left[ \fp{}^{fa}{}_{[\ti\tj]} + \frac{\bar{\eta}}{2(4!)^2}
\e_{\ti\tj\tf_1...\tf_6} \left(T^{[\tf_1\tf_2\tf_3]}\bs^{fa}
T^{[\tf_4\tf_5\tf_6]} \right) \right]
\nn\\
 & & - \frac{1}{4(4!)} \left(T^{[\tf\tb\ta]} \bs^f \s^a
T^{[\ti\tj\tk]} \right) \OM_f{}_{[\tf\ti\tj\tk]}
+ \frac{\vi}{16} \left(T^{[\tf\tb\ta]} \bs^{fg} \bs^a
T_{[\tf\ti\tj]} \right) \fm_{fg}{}^{[\ti\tj]}
\nn\\
 & & + \frac{\vi \eta}{2(4!)^3} \e^{\ti\tj\tk\tl\tm\te\tf\tg}
\left( T_{[\td\tl\tm]}T_{[\te\tf\tg]} \right) \left(T^{[\td\tb\ta]}
\bs^a T_{[\ti\tj\tk]} \right) \,.
\label{em_vect_u}
\eea
Due to the self--duality properties of these field strengths, these equations of motion are related to
their Bianchi identities as follows
\be
\frac{\vi}{2}\e^{facb}\cd_f\fp_{cb}{}_{[\tb\ta]}\ =\ \cd_f\fp{}^{fa}{}_{[\tb\ta]}\,,
\qquad\frac{\vi}{2}\e^{facb}\cd_f\fm_{cb}{}^{[\tb\ta]}\ =\ -\cd_f\fm{}^{fa}{}^{[\tb\ta]}\,.
\ee
Again, it is worthwhile to notice that the Bianchi identities of the graviphotons can be found directly
among the super Bianchi identities, namely they are the identities $\brg{_{dcb}{}^\au}$. As a matter of
fact the form given here can be recovered from the Bianchi identity $\brg{_{dcb}{}^\au}$ after converting
the central charge index into $SU(8)$ ones by multiplying by the scalar matrices $S_\au{}^{[\tb\ta]}$ and
$S_\au{}_{[\tb\ta]}$, as well as using the equations of motion for the gravitini fields presented below
in equations \equ{em_gravitini1} and \equ{em_gravitini2}.

On the other hand, take the spinorial derivative $\cd^\dd_\td$ of the same Dirac equation \equ{em_spin_u}
and after commuting on the left--hand--side this derivative with that of the space--time one, take the
totally antisymmetric part in the $SU(8)$ indices. Now, the antisymmetric part in the spinorial indices
$\dd$ and $\da$ gives the equations of motion for the scalars written for the field strength component
$\OM_f{}_{[\td\tc\tb\ta]}$
\bea
\cd ^f \OM_f{}_{[\td\tc\tb\ta]} & = &
-2\d_{\td\tc\tb\ta}^{\te_1...\te_4}\fp{}^{ba}{}_{[\te_1\te_2]} \fp_{ba}{}_{[\te_3\te_4]}
- 2 \bar{\eta}\e_{\td\tc\tb\ta\te_1...\te_4} \fm{}^{ba}{}^{[\te_1\te_2]} \fm_{ba}{}^{[\te_3\te_4]}\nn \\
 & & + \frac{\vi}{2 (3!) (4!)}\d_{\td\tc\tb\ta}^{\te_1...\te_4}\OM_f{}_{[\te_1\te_2\te_3\te_4]}
\left(T_{[\tf_1\tf_2\tf_3]} \s^f T^{[\tf_1\tf_2\tf_3]} \right) \nn \\
 & & - \frac{\vi}{2 (4!)}\d_{\td\tc\tb\ta}^{\te_1...\te_4}\OM_f{}_{[\te_1\te_2\te_3\tf_1]}
\left(T_{[\te_4\tf_2\tf_3]} \s^f T^{[\tf_1\tf_2\tf_3]} \right) \nn \\
 & & - \frac{\vi}{32}\d_{\td\tc\tb\ta}^{\te_1...\te_4}\OM_f{}_{[\te_1\te_2\tf_1\tf_2]}
\left(T_{[\te_3\te_4\tf_3]}\s^f T^{[\tf_1\tf_2\tf_3]} \right) \nn \\
 & & + \frac{\vi}{2 (3!)^2}\d_{\td\tc\tb\ta}^{\te_1...\te_4}\OM_f{}_{[\te_1\tf_1\tf_2\tf_3]}
\left(T_{[\te_2\te_3\te_4]} \s^f T^{[\tf_1\tf_2\tf_3]} \right) \nn \\
 & & + \frac{5 \eta}{2 (3!) (4!)^3}\d_{\td\tc\tb\ta}^{\te_1...\te_4}\e^{\tf_1 \ldots \tf_8}
\left(T_{[\te_1\te_2\te_3]}T_{[\tf_1\tf_2\tf_3]}\right)
\left(T_{[\te_4\tf_4\tf_5]}T_{[\tf_6\tf_7\tf_8]}\right) \nn \\
 & & - \frac{\eta}{256 (4!)}\d_{\td\tc\tb\ta}^{\te_1...\te_4}\e^{\tf_1 \ldots \tf_8}
\left(T_{[\te_1\te_2\tf_1]}T_{[\te_3\tf_2\tf_3]}\right)
\left(T_{[\te_4\tf_4\tf_5]}T_{[\tf_6\tf_7\tf_8]}\right) \nn \\
 & & + \frac{5 \bar{\eta}^2}{2 (3!) (4!)^3}
\e_{\td\tc\tb\ta\te_1\ldots\te_4} \e_{\tf_1 \ldots \tf_8}
\left(T^{[\te_1\te_2\te_3]}T^{[\tf_1\tf_2\tf_3]}\right)
\left(T^{[\te_4\tf_4\tf_5]}T^{[\tf_6\tf_7\tf_8]}\right) \nn \\
 & & - \frac{\bar{\eta}^2}{256 (4!)}
\e_{\td\tc\tb\ta\te_1\ldots\te_4} \e_{\tf_1 \ldots \tf_8}
\left(T^{[\te_1\te_2\tf_1]}T^{[\te_3\tf_2\tf_3]}\right)
\left(T^{[\te_4\tf_4\tf_5]}T^{[\tf_6\tf_7\tf_8]}\right).
\label{em_scalar}
\eea
The equations of motion for the field strength component $\OM_f{}^{[\td\tc\tb\ta]}$,
\bea
\cd ^f \OM_f{}^{[\td\tc\tb\ta]}&=&
-2\d^{\td\tc\tb\ta}_{\te_1...\te_4}\fm{}^{ba}{}^{[\te_1\te_2]}\fm_{ba}{}^{[\te_3\te_4]}
-2\eta\e^{\td\tc\tb\ta\te_1...\te_4}  \fp{}^{ba}{}_{[\te_1\te_2]}\fp_{ba}{}_{[\te_3\te_4]}\nn \\
 & & - \frac{\vi}{2 (3!) (4!)}\d^{\td\tc\tb\ta}_{\te_1...\te_4}\OM_f{}^{[\te_1\te_2\te_3\te_4]}
\left(T_{[\tf_1\tf_2\tf_3]} \s^f T^{[\tf_1\tf_2\tf_3]} \right) \nn \\
 & & + \frac{\vi}{2 (4!)}\d^{\td\tc\tb\ta}_{\te_1...\te_4}\OM_f{}^{[\te_1\te_2\te_3\tf_1]}
\left(T_{[\tf_1\tf_2\tf_3]} \s^f T^{[\te_4\tf_2\tf_3]} \right) \nn \\
 & & + \frac{\vi}{32}\d^{\td\tc\tb\ta}_{\te_1...\te_4}\OM_f{}^{[\te_1\te_2\tf_1\tf_2]}
\left(T_{[\tf_1\tf_2\tf_3]} \s^f T^{[\te_3\te_4\tf_3]} \right) \nn \\
 & & - \frac{\vi}{2 (3!)^2} \d^{\td\tc\tb\ta}_{\te_1...\te_4}\OM_f{}^{[\te_1\tf_1\tf_2\tf_3]}
\left(T_{[\tf_1\tf_2\tf_3]} \s^f T^{[\te_2\te_3\te_4]} \right) \nn \\
 & & + \frac{5 \bar{\eta}}{2 (3!) (4!)^3} \d^{\td\tc\tb\ta}_{\te_1...\te_4}\e_{\tf_1 \ldots \tf_8}
\left(T^{[\te_1\te_2\te_3]}T^{[\tf_1\tf_2\tf_3]}\right)
\left(T^{[\te_4\tf_4\tf_5]}T^{[\tf_6\tf_7\tf_8]}\right) \nn \\
 & & - \frac{\bar{\eta}}{256 (4!)} \d^{\td\tc\tb\ta}_{\te_1...\te_4} \e_{\tf_1 \ldots \tf_8}
\left(T^{[\te_1\te_2\tf_1]}T^{[\te_3\tf_2\tf_3]}\right)
\left(T^{[\te_4\tf_4\tf_5]}T^{[\tf_6\tf_7\tf_8]}\right) \nn \\
 & & + \frac{5 \eta^2}{2 (3!) (4!)^3}
\e^{\td\tc\tb\ta\te_1\ldots\te_4} \e^{\tf_1 \ldots \tf_8}
\left(T_{[\te_1\te_2\te_3]}T_{[\tf_1\tf_2\tf_3]}\right)
\left(T_{[\te_4\tf_4\tf_5]}T_{[\tf_6\tf_7\tf_8]}\right) \nn \\
 & & - \frac{\eta^2}{256 (4!)}
\e^{\td\tc\tb\ta\te_1\ldots\te_4} \e^{\tf_1 \ldots \tf_8}
\left(T_{[\te_1\te_2\tf_1]}T_{[\te_3\tf_2\tf_3]}\right)
\left(T_{[\te_4\tf_4\tf_5]}T_{[\tf_6\tf_7\tf_8]}\right),
\eea
can be calculated in an analogous way from the Dirac equation \equ{em_spin_d} and one can check that of
course, the two equations are related by $\eta$-duality relations.

The symmetric part in the indices $\dd$ and $\da$ is the anti--self--dual part of the Bianchi identity for
the scalars,
\be
\cd_d \OM_c^{[\td\tc\tb\ta]} - \cd_c \OM_d^{[\td\tc\tb\ta]}\ =\
-32\left[\left(T_{dc}{}^{[\td} T^{[\tc\tb\ta]]}\right)
+\frac{\eta}{4!} \eps^{\td\tc\tb\ta\tf\ti\tj\tk} \left(T_{dc}{}_\tf T_{[\ti\tj\tk]}\right)\right],
\ee
which can be deduced also from the off--diagonal part of the superspace Bianchi identities
\equ{DOM} of the one--form $\OM$.

Unlike the equations of motion presented above, in Poincar\'e supergravity the equations of motion for the
gravitini and graviton cannot be obtained by acting again by spinorial derivatives on the equations of
motion obtained so far. One could obtain this way only the Bianchi identities for these fields (which can
be obtained also in a much direct way: the Bianchi identities of the gravitini are exactly the superspace
Bianchi identities $\brg{_{dcb}{}^\a_\ta}$ and $\brg{_{dcb}{}^\ta_\da}$, since the Bianchi identity for
the graviton is just the $dcb$ component of the second Bianchi identity \equ{idbR} written for the
Lorentz curvature). The equations of motion for the gravitini and of the graviton are directly given by
the superspace Bianchi identities.

For example, the Bianchi identities $\brg{_\d^\td{}_c{}{}^\tb_\b{}^\a_\ta}$ and
$\brg{^\dd_\td{}_c{}_\tb^\db{}_\da^\ta}$ give the irreducible components $T_{(\b\a)}{}_\ta^\a$ and
$T^{(\db\da)}{}_\ta^\a$ of the gravitini field strength $T_{ba}{}_\ta^\a$ and respectively the
irreducible components $T_{(\b\a)}{}^\ta_\da$ and $T^{(\db\da)}{}^\ta_\da$ of the gravitini field strength
$T_{ba}{}^\ta_\da$ as a non--linear function of basic superfields. It turns out that the knowledge of
these parts is sufficient to write down the equations of motion of the gravitini fields:
\bea
\e^{dcba}(\bs_c T_{ba}{}_\ta)^\da&=&\frac{1}{3!}
(\bs^a\s^dT^{[\td\tc\tb]})^\da\OM_a{}_{[\td\tc\tb\ta]}
+\vi(\bs^{ba}\bs^dT_{[\tc\tb\ta]})^\da \cf_{ba}^-{}^{[\tc\tb]}\nn\\[2mm]
&&-\frac{\vi\eta}{3(4!)^2}\,\e^{\tf_1...\tf_8}
(T_{[\ta\tf_1\tf_2]}T_{[\tf_3\tf_4\tf_5]})(\bs^dT_{[\tf_6\tf_7\tf_8]})^\da\,,
\label{em_gravitini1}
\eea
\bea
\e^{dcba}(\s_c T_{ba}{}^\ta)_\a&=&-\frac{1}{3!}
(\s^a\bs^dT_{[\td\tc\tb]})_\a\OM_a{}^{[\td\tc\tb\ta]}
-\vi(\s^{ba}\s^dT^{[\tc\tb\ta]})_\a \cf_{ba}^+{}_{[\tc\tb]}\nn\\[2mm]
&&+\frac{\vi\eta}{3(4!)^2}\,\e_{\tf_1...\tf_8}
(T^{[\ta\tf_1\tf_2]}T^{[\tf_3\tf_4\tf_5]})(\s^dT^{[\tf_6\tf_7\tf_8]})_\a\,.
\label{em_gravitini2}
\eea

Finally, in order to give the equations of motion for the graviton we need the expression of the
supercovariant Ricci tensor, $R_{db}=R_{dcba}\eta^{ca}$, which is given \equ{Riccitensor} by the
superspace Bianchi identities at canonical dimension 2. The corresponding Ricci scalar,
$R=R_{db}\eta^{db}$, is then also determined \equ{Ricciscalar} and using the equations of motion for the
spinor fields we find that the Einstein equation takes the form
\bea
R_{db}-\frac{1}{2}\eta_{db}R
&=&-\frac{1}{3!}\left[\OM_{d}{}^{[\ti\tj\tk\tl]} \OM_{b[\ti\tj\tk\tl]}
-\frac{1}{2} \eta_{db} \OM^{f[\ti\tj\tk\tl]}\OM_{f[\ti\tj\tk\tl]} \right]
-32 \fp_{(d}{}^f{}_{[\ti\tj]} \fm_{b)f}{}^{[\ti\tj]} \nn \\
& & - \frac{\vi}{4!} \left[ \left(T_{[\ti\tj\tk]} \s_{(d}
\cd_{b)} T^{[\ti\tj\tk]}\right) + \left(T^{[\ti\tj\tk]}
\bs_{(d} \cd_{b)} T_{[\ti\tj\tk]}\right) \right] \nn \\
 & &+\frac{\eta}{3!4!} \e^{\tg_1\ldots\tg_8}
\left[\fp_{(d}{}^f{}_{[\tg_1\tg_2]} \left(T_{[\tg_3\tg_4\tg_5]}
\s_{b)f} T_{[\tg_6\tg_7\tg_8]} \right)\right. \nn \\
 & & \phantom{+\frac{\eta}{3!4!} \e^{\tg_1\ldots\tg_8}[}\left.- \frac{5}{4} \eta_{db} \fp{}^{ef}{}_{[\tg_1\tg_2]} \left(T_{[\tg_3\tg_4\tg_5]}
\s_{ef} T_{[\tg_6\tg_7\tg_8]} \right)\right] \nn \\
& &+\frac{\bar{\eta}}{3!4!} \e_{\tg_1\ldots\tg_8}
\left[\fm_{(d}{}^f{}^{[\tg_1\tg_2]} \left(T^{[\tg_3\tg_4\tg_5]}
\bs_{b)f} T^{[\tg_6\tg_7\tg_8]} \right)\right. \nn \\
 & & \phantom{+\frac{\eta}{3!4!} \e^{\tg_1\ldots\tg_8}[}\left.- \frac{5}{4} \eta_{db} \fm{}^{ef}{}^{[\tg_1\tg_2]} \left(T_{[\tg_3\tg_4\tg_5]}
\bs_{ef} T^{[\tg_6\tg_7\tg_8]} \right)\right] \nn \\
 & & + \frac{1}{64( 4!)^2} \d_{\th_1}^{\tg_1}{}_{\ldots}^{\ldots}
{}_{\th_6}^{\tg_6} \left( T_{[\tg_1\tg_2\tg_3]} \s_{d} T^{[\th_1\th_2\th_3]}
\right) \left( T_{[\tg_4\tg_5\tg_6]} \s_{b} T^{[\th_4\th_5\th_6]}
\right) \nn \\
 & & - \frac{1}{32 (4!)^2} \eta_{db} \d_{\th_1}^{\tg_1}{}
_{\ldots}^{\ldots}{}_{\th_5}^{\tg_5} \left( T_{[\tf\tg_1\tg_2]} T_{[\tg_3\tg_4\tg_5]} \right) \left(
T^{[\tf\th_1\th_2]}
T^{[\th_3\th_4\th_5]} \right) \nn \\
& & + \frac{1}{4 (4!)^2} \left[ \left(T_{[\ti\tj\tk]}
\s_{d} T^{[\ti\tj\tk]} \right) \left(T_{[\tl\tm\tn]} \s_{b}
T^{[\tl\tm\tn]}\right) \right. \nn \\
 & & \phantom{+ \frac{1}{4 \times (4!)^2}[}  \left.
- 9 \left(T_{[\tg\ti\tj]} \s_{d}  T^{[\tf\ti\tj]} \right)
\left(T_{[\tf\tk\tl]} \s_{b} T^{[\tg\tk\tl]} \right) \right]\nn \\
 & & - \frac{1}{(4!)^2} \eta_{db} \left(T_{[\ti\tj\tk]}
T_{[\tl\tm\tn]}\right) \left[ \left(T^{[\ti\tj\tk]} T^{[\tl\tm\tn]}
\right)+\frac{27}{4}\left(T^{[\ti\tj\tl]}T^{[\tk\tm\tn]}\right)\right]\,,
\label{gr}
\eea
where one may recognize in the first two lines of the right--hand--side the usual terms of the
energy--momentum tensor corresponding to matter fields: scalar fields, photon fields and spinor fields
respectively. The contribution of the gravitini is hidden in the left--hand--side, in the component
development \cite{GK02} of $R_{db}$.

The Einstein equation completes the ensemble of the equations of motion for the component fields.
Fortunately, in the article \cite{dWN82}, which contains a detailed list of the results obtained in the
component formulation, most of the equations of motion are given in terms of super--covariant quantities,
given in paragraph \ref{sup_comp} as functions of the component fields. As a consequence, it is easy to
compare the equations of motion deduced in this paragraph from the geometry with the component results of
\cite{dWN82}, and see that there is a perfect concordance between them.

%%%%%%%%%%%%%%%%%%%%%%%%%%%%%%%%%%%%%%%%%%%%%%%%%%%%%%%
\section{Conclusion}
%%%%%%%%%%%%%%%%%%%%%%%%%%%%%%%%%%%%%%%%%%%%%%%%%%%%%%%

We have presented here a new approach to the superspace formulation of $N=8$ supergravity, using central charge superspace. The presence
of the central charge coordinates is essential in the formulation. It permits to identify the gauge
vectors of the theory in the super--vielbein on the same footing with the graviton and gravitini, and
also, it allows to identify the scalars directly, as lowest canonical dimension torsion components. In
addition, we recover the well--known essential properties of the multiplet as consequences of the
geometric structure: we deduce the self--duality properties of the vectors as well as the
$E_{7(+7)}/SU(8)$ non--linear $\sigma$ model structure for the scalars using straightforward superspace
techniques in central charge superspace.

It is worthwhile to note here that there exists a formal correspondence between the formulation in
central charge superspace presented here and the geometric formulation in superspace extended by $56$
bosonic coordinates \cite{HL81} presented in detail by Howe \cite{How82}. The correspondence, described
 in more detail in \cite{Kis00}, is based on a simple redefinition of the frame of the type
\be
\tilde{E}\ =\ E\cx\,,\quad\textrm{with}\quad
\cx=\left(\begin{array}{cc}
\d^{\underline{\ca}}_{\underline{\cb}}& 0\\
0 &S
\end{array}\right),
\eqn{Howe}
that is, on a rotation by the scalars $S$ in the central charge sector\footnote{Underlined indices denote
the ordinary superspace sector $\underline{\ca}=(a,\,{}_\a^\ta,\,{}^\da_\ta)$.}. However, the difference
between the two approaches is conceptual in the sense that in Howe's approach the translation generators
in the extra bosonic coordinates are not "genuine" central charges, they transform under $SU(8)$. Indeed, the frame
\equ{Howe} in the extra bosonic sector $(\tilde{E}_{[\tb\ta]},\,\tilde{E}^{[\tb\ta]})$ and thus also the
translation generators in the direction of the extra bosonic coordinates carry $SU(8)$ representation
indices.

The identification of the $N=8$ supergravity in the geometry of central charge superspace described here
is complete, since we also presented in detail the deduction from the geometry of both the supergravity
transformation laws and the equations of motion for the component fields. Our results obtained from the
geometric formulation are in perfect accord with the results in terms of super--covariant quantities
obtained in component formulation given in the article \cite{dWN82}.

%%%%%%%%%%%%%%%%%%%%%%%%%%%%%%%%%%%%%%%%%%%%%%%%%%%%%%%
\section*{Acknowledgements}
%%%%%%%%%%%%%%%%%%%%%%%%%%%%%%%%%%%%%%%%%%%%%%%%%%%%%%%

We thank Richard Grimm for the numerous discussions, for his support, as well as for inciting us to write
this article. We also appreciate remarks of Warren Siegel and Paul Howe on the preliminary version of this
article. We finally would like to thank for the hospitality of the Center of Theoretical Physics of
Marseille, where this work was achieved.

\begin{appendix}

%%%%%%%%%%%%%%%%%%%%%%%%%%%%%%%%%%%%%%%%%%%%%%%%%%%%%%%
\section{Totally antisymmetric $U(N)$ tensors and deltas}
%%%%%%%%%%%%%%%%%%%%%%%%%%%%%%%%%%%%%%%%%%%%%%%%%%%%%%%
\label{deltas}

In this appendix we collect some useful relations concerning totally antisymmetric tensors and deltas in $U(N)$ 
indices.

$\d^{\ta}_\tb$ being the Kronecker symbol, we can define the objects with antisymmetric indices
\bea
\d^{\ta_1\ta_2}_{\tb_1\tb_2}&=&\d^{\ta_1}_{\tb_1}\d^{\ta_2}_{\tb_2}-\d^{\ta_2}_{\tb_1}\d^{\ta_1}_{\tb_2}\nn\\
\d^{\ta_1\ta_2\ta_3}_{\tb_1\tb_2\tb_3}&=&\d^{\ta_1}_{\tb_1}\d^{\ta_2\ta_3}_{\tb_2\tb_3}
+\d^{\ta_2}_{\tb_1}\d^{\ta_3\ta_1}_{\tb_2\tb_3}+\d^{\ta_3}_{\tb_1}\d^{\ta_1\ta_2}_{\tb_2\tb_3}\nn, \quad etc.
\eea
In general, for $k$ antisymmetric indices we define the corresponding objects by recursion
\bea
\d^{\ta_1 ...\ta_k}_{\tb_1 ...\tb_k}&=&\sum^k_{i=1}(-1)^{(k-1)(i+1)}\d^{\ta_i}_{\tb_1}
\d^{\ta_{i+1}...\ta_k\ta_1...\ta_{i-1}}_{\tb_2\qquad...\qquad\tb_k}\nn\\
&=&\sum^k_{i=1}(-1)^{(k-1)(i+1)}\d^{\ta_1}_{\tb_i}\d^{\ta_2\qquad...\qquad\ta_k}_{
\tb_{i+1}...\tb_k\tb_1...\tb_{i-1}}
\eea
and call them "generalized deltas". These generalized deltas have the property
\bea
\d^{\ta_1 ...\ta_k}_{\tb_1 ...\tb_k}\ X^{[\tb_1 ...\tb_k]}&=&k!\ X^{[\ta_1 ...\ta_k]}\ ,
\qquad\forall X^{[\ta_1 ...\ta_k]}\,, \\
\d_{\ta_1 ...\ta_k}^{\tb_1 ...\tb_k}\ X_{[\tb_1 ...\tb_k]}&=&k!\ X_{[\ta_1 ...\ta_k]}\ ,
\qquad\forall X_{[\ta_1 ...\ta_k]}\,,
\eqan{unite}
thus, the operator $\frac{1}{k!}\d^{\ta_1 ...\ta_k}_{\tb_1 ...\tb_k}$ acts as the unit operator in the
space of tensors which are antisymmetric in $k$ indices. If we contract an upper index with a lower one
in a generalized delta we have
\be
\d^{\ta_1\ta_2 ...\ta_k}_{\ta_1\tb_2 ...\tb_k}\ =\ (N-k+1)\d^{\ta_2 ...\ta_k}_{\tb_2 ...\tb_k}\,,
\eqn{N-k+1}
as for the contraction of $l$ indices, this implies
\be
\d^{\ta_1...\ta_l\ta_{l+1}...\ta_k}_{\ta_1...\ta_l\tb_{l+1} ...\tb_k}
\ =\ (N-k+1)(N-k+2)...(N-k+l)\d^{\ta_{l+1} ...\ta_k}_{\tb_{l+1} ...\tb_k}.
\eqn{N-k+l}

Let $\e^{\ta_1\ta_2 ...\ta_N}$ be the totally antisymmetric tensor. Choosing $\e^{1...N}=1$ and
$\e_{1...N}=1$ we have the relation
\be
\e^{\ta_1 ...\ta_N}\e_{\tb_1 ...\tb_N}\ =\ \d^{\ta_1 ...\ta_N}_{\tb_1 ...\tb_N}.
\eqn{epseps}
Then, if we contract $k$ indices on a product of two totally antisymmetric tensors $\e$, we obtain as
consequence of \equ{epseps} and of the property \equ{N-k+l}
\be
\e^{\ta_1...\ta_k\ta_{k+1} ...\ta_N}\e_{\ta_1...\ta_k\tb_{k+1} ...\tb_N}
\ =\ k!\ \d^{\ta_{k+1} ...\ta_N}_{\tb_{k+1} ...\tb_N}
\ee
Finally, since the property \equ{unite} is valid also for $\e$, we have
\be
\d^{\ta_1 ...\ta_k}_{\tb_1 ...\tb_k}\ \e^{\tb_1 ...\tb_k\ta_{k+1} ...\ta_N}
\ =\ k!\ \e^{\ta_1 ...\ta_k\ta_{k+1} ...\ta_N}\,.
\ee

%%%%%%%%%%%%%%%%%%%%%%%%%%%%%%%%%%%%%%%%%%%%%%%%%%%%%%%
\section{Solution of the Bianchi identities}
%%%%%%%%%%%%%%%%%%%%%%%%%%%%%%%%%%%%%%%%%%%%%%%%%%%%%%%
\label{sol}

In this appendix we give the solution of the Bianchi identities, which is compatible with the constraints
presented in section \ref{id} and corresponds to the $N=8$ supergravity in central charge superspace.
This means  that all torsion and curvature components are expressed as polynomial functions of the scalar
superfields $T^{[\tc\tb]}{}^\au$ and $T_{[\tc\tb]}{}^\au$ (or their super--covariant field strength
$\OM_a$), the gravigini superfields $T_{[\tc\tb\ta]\a}$ and $T^{[\tc\tb\ta]\da}$, the gravitini Weyl
tensors $\S_{(\g\b\a)\ta}$ and $\S^{(\dg\db\da)\ta}$, the usual Weyl tensors $V_{(\d\g\b\a)}$ and
$V_{(\d\g\b\a)}$, as well as the super--covariant field strength of the graviphotons, identified in the
torsion component $T_{cb}{}^\au$, denoted also $F_{cb}{}^\au$. Actually, as we already have seen in
section \ref{selfdphoton}, the dynamical parts of this field strengths of graviphotons are only the fields
$\fp_{cb}{}_{[\tb\ta]}$ and $\fm_{cb}{}^{[\tb\ta]}$, which are subject to the self--duality relations
\equ{self}, and only these parts appear in torsion and curvature components.
Moreover, the spinorial derivatives of these basic superfields are also expressed as polynomial functions
of themselves.

In this appendix we will give all non--vanishing torsion and curvature components as well as the
spinorial covariant derivatives of the basic superfields listed above.

\subsection*{Non--vanishing torsion components}

The list of the non--vanishing torsion components is the following:
\be
 T{^\tc_\g}{^\db_\tb}{^a}\ =\
-2\vi\d{^\tc_\tb}(\s{^a}\eps){_\g}{^\db}~, \qquad
T{^\tc_\g}{^\tb_\b}{^\au}\ =\ \eps_{\g\b}T^{[\tc\tb]\au}~,\qquad T{^\dg_\tc}{^\db_\tb}{^\au}\ =\
\eps^{\dg\db}T{_{[\tc\tb]}}{}^\au~,
\ee
\be
\begin{array}{rclcrcl}
T^\tc_\g{}^\tb_\b{}^\ta_\da& =& \eps_{\b\g}T^{[\tc\tb\ta]}{}_\da,&
\qquad &T_\tc^\dg{}_\tb^\db{}_\ta^\a& =&\eps_{\db\dg}T_{[\tc\tb\ta]}{}^\a\,,
\\[5mm]
T^\tc_\g{}_b{}^\au& =& -\frac{\vi}{8}(\s_bT^{[\tc\ti\tj]})_\g T_{[\ti\tj]}{}^\au\,,&
\qquad& T_\tc^\dg{}_b{}^\au& =& -\frac{\vi}{8}(\bs_bT_{[\tc\ti\tj]})^\dg T^{[\ti\tj]\au}\,,
\\[5mm]
T_\g^\tc{}_b{}_\da^\ta& =& 4\vi(\s^f)_{\g\da}F_{fb}{}^{[\tc\ta]}\,,
&\qquad& T_\tc^\dg{}_b{}_\ta^\a& =& 4\vi(\bs^f)^{\dg \a}F_{fb[\tc\ta]}\,,
\\[5mm]
T_\g^\tc{}_b{}_\ta^\a &=& -\d_\g^\a T_b^\tc{}_\ta-2(\s_{bf})_\g{}^\a U^{f\tc}{}_\ta\,,&\qquad&
T_\tc^\dg{}_b{}_\da^\ta &=& \d_\da^\dg T_b^\ta{}_\tc+2(\bs_{bf})^\dg{}_\da U^{f\ta}{}_\tc\,,
\end{array}
\ee
\bea
T_c{}_b{}_{\ta\a}&=&-(\eps\s_{cb})^{\g\b}\S_{(\g\b\a)\ta}
+\frac{\vi}{18}(\d^{fg}_{cb}-\frac{\vi}{2}\e_{cb}{}^{fg})\OM_f{}_{[\ta\tb\tc\td]}
(\s_gT^{[\tb\tc\td]})_\a\nn\\[2mm]
&&+\fm_{cb}{}^{[\tc\tb]}T_{[\tc\tb\ta]\a}
-\frac{\eta}{4!(3!)^3}\e^{\tf_1...\tf_6\tc\tb}(\s_{cb}T_{[\tf_1\tf_2\tf_3]})_\a
(T_{[\tf_4\tf_5\tf_6]}T_{[\tc\tb\ta]})\,,\\[2mm]
T_c{}_b{}^{\ta\da}&=&-(\eps\bs_{cb})_{\dg\db}\S^{(\dg\db\da)\ta}
+\frac{\vi}{18}(\d^{fg}_{cb}+\frac{\vi}{2}\e_{cb}{}^{fg})\OM_f{}^{[\ta\tb\tc\td]}
(\bs_gT_{[\tb\tc\td]})^\da\nn\\[2mm]
&&+\fp_{cb}{}_{[\tc\tb]}T^{[\tc\tb\ta]\da}
-\frac{{\bar{\eta}}}{4!(3!)^3}\e_{\tf_1...\tf_6\tc\tb}(\bs_{cb}T^{[\tf_1\tf_2\tf_3]})^\da
(T^{[\tf_4\tf_5\tf_6]}T^{[\tc\tb\ta]})\,,
\eea
where the superfields $T_b{}^\tb{}_\ta$ and $U_b{}^\tb{}_\ta$ are quadratic terms in the spinor
superfields
\be
T_b{}^\tb{}_\ta\ =\ -\frac{\vi}{16}T_{[\ta\ti\tj]}\s_bT^{[\tb\ti\tj]}\,,
\qquad U_b{}^\tb{}_\ta\ =\ \frac{\vi}{16}\left(T_{[\ta\ti\tj]}\s_bT^{[\tb\ti\tj]}
-\frac{1}{6}\d_\ta^\tb T_{[\ti\tj\tk]}\s_bT^{[\ti\tj\tk]}\right)\,,
\ee
while for the field strength components $F_{ba}{}_{[\tb\ta]}$ and  $F_{ba}{}^{[\tb\ta]}$ we have
\bea
F_{ba}{}_{[\tb\ta]}&=&\fp_{ba}{}_{[\tb\ta]}
-\frac{\eta}{2(4!)^2}\e_{\tb\ta\tf_1...\tf_6}T^{[\tf_1\tf_2\tf_3]}\bs_{ba}T^{[\tf_4\tf_5\tf_6]}\,,\\[2mm]
F_{ba}{}^{[\tb\ta]}&=&\fm_{ba}{}^{[\tb\ta]}
-\frac{\bar{\eta}}{2(4!)^2}\e^{\tb\ta\tf_1...\tf_6}T_{[\tf_1\tf_2\tf_3]}\s_{ba}T_{[\tf_4\tf_5\tf_6]}\,.
\eea

%{\it Juste pour nous: $\S=\frac{1}{3}W$}

\subsection*{Curvature components}

Let us begin with giving the list of Lorentz curvature components:
\be
R_\d^\td{}_\g^\tc{}_{ba} =\ -16 \eps_{\d\g} F_{ba}{}^{[\td\tc]}\,,
\quad R_\td^\dd{}_\tc^\dg{}_{ba} =\ -16\eps^{\dd\dg} F_{ba}{}_{[\td\tc]}\,,
\quad R_\d^\td{}_\tc^\dg{}_{ba} =\ 4 \e_{badc}(\s^d\eps)_\d{}^\dg U^{c\td}{}_\tc \,.
\ee
\bea
R^\td_\d{}_c{}_b{}_a&=&\vi\left(\s_cT_{ba}{}^\td-\s_aT_{cb}{}^\td-\s_bT_{ac}{}^\td\right)_\d\,,\\[2mm]
R_\td^\dd{}_c{}_b{}_a&=&\vi\left(\bs_cT_{ba}{}_\td-\bs_aT_{cb}{}_\td-\bs_bT_{ac}{}_\td\right)^\dd\,,
\eea
\bea
R_{dc}{}_{ba}&=&(\eps\s_{dc})^{\d\g}(\eps\s_{ba})^{\b\a}V_{(\d\g\b\a)}
+(\eps\bs_{dc})_{\dd\dg}(\eps\bs_{ba})_{\db\da}V^{(\dd\dg\db\da)}\nn\\[2mm]
&&+\frac{1}{2}\left(\eta_{db}R_{ca}-\eta_{da}R_{cb}+\eta_{ca}R_{db}-\eta_{cb}R_{da}\right)
-\frac{1}{6}(\eta_{db}\eta_{ca}-\eta_{da}\eta_{cb})R
\eea
with the supercovariant Ricci tensor, $R_{db}=R_{dcba}\eta^{ca}$, given by
\bea
R_{db}\  &=& - \frac{1}{3!}\OM_{d}{}^{[\ti\tj\tk\tl]}\OM_{b[\ti\tj\tk\tl]}
-32 \fp_{(d}{}^f{}_{[\ti\tj]} \fm_{b)f}{}^{[\ti\tj]} \nn \\
 & &+\frac{\eta}{3!4!} \e^{\tg_1\ldots\tg_8}
\left[\fp_{(d}{}^f{}_{[\tg_1\tg_2]} \left(T_{[\tg_3\tg_4\tg_5]}
\s_{b)f} T_{[\tg_6\tg_7\tg_8]} \right)\right. \nn \\
 & & \phantom{+\frac{\eta}{3!4!} \e^{\tg_1\ldots\tg_8}[}\left.+ \frac{1}{4} \eta_{db} \fp{}^{ef}{}_{[\tg_1\tg_2]} \left(T_{[\tg_3\tg_4\tg_5]}
\s_{ef} T_{[\tg_6\tg_7\tg_8]} \right)\right] \nn \\
& &+\frac{\bar{\eta}}{3!4!} \e_{\tg_1\ldots\tg_8}
\left[\fm_{(d}{}^f{}^{[\tg_1\tg_2]} \left(T^{[\tg_3\tg_4\tg_5]}
\bs_{b)f} T^{[\tg_6\tg_7\tg_8]} \right)\right. \nn \\
 & & \phantom{+\frac{\eta}{3!4!} \e^{\tg_1\ldots\tg_8}[}\left.+ \frac{1}{4} \eta_{db} \fm{}^{ef}{}^{[\tg_1\tg_2]} \left(T_{[\tg_3\tg_4\tg_5]}
\bs_{ef} T^{[\tg_6\tg_7\tg_8]} \right)\right] \nn \\
& & -\frac{\vi}{4!} \left[ \left( T_{[\ti\tj\tk]} \s_{(d} \cd_{b)} T^{[\ti\tj\tk]} \right) - \frac{1}{4}
\eta_{db} \left( T_{[\ti\tj\tk]}
\s^f \cd_f T^{[\ti\tj\tk]} \right) \right. \nn \\
 & & \phantom{- \frac{\vi}{4!} [} \left. + \left( T^{[\ti\tj\tk]} \bs
_{(d} \cd_{b)} T_{[\ti\tj\tk]} \right) - \frac{1}{4} \eta_{db} \left(
T^{[\ti\tj\tk]} \bs^f \cd_f T_{[\ti\tj\tk]} \right) \right] \nn \\
 & & + \frac{1}{64( 4!)^2} \d_{\th_1}^{\tg_1}{}_{\ldots}^{\ldots}
{}_{\th_6}^{\tg_6} \left( T_{[\tg_1\tg_2\tg_3]} \s_{d} T^{[\th_1\th_2\th_3]}
\right) \left( T_{[\tg_4\tg_5\tg_6]} \s_{b} T^{[\th_4\th_5\th_6]}
\right) \nn \\
 & & + \frac{1}{32 (4!)^2} \eta_{db} \d_{\th_1}^{\tg_1}{}
_{\ldots}^{\ldots}{}_{\th_5}^{\tg_5} \left( T_{[\tf\tg_1\tg_2]} T_{[\tg_3\tg_4\tg_5]} \right) \left(
T^{[\tf\th_1\th_2]}
T^{[\th_3\th_4\th_5]} \right) \nn \\
& & + \frac{1}{4  (4!)^2} \left[ \left(T_{[\ti\tj\tk]} \s_{d}T^{[\ti\tj\tk]} \right)
\left(T_{[\tl\tm\tn]} \s_{b} T^{[\tl\tm\tn]}\right) \right. \nn \\
& & \phantom{+ \frac{1}{4 \times (4!)^2} [} \left. -9\left(T_{[\tg\ti\tj]}
\s_{d} T^{[\tf\ti\tj]} \right) \left(T_{[\tf\tl\tm]} \s_{b}T^{[\tg\tl\tm]}\right) \right]\nn \\
&&+\frac{1}{2(4!)^2}\eta_{db}\left(T_{[\ti\tj\tk]}T_{[\tl\tm\tn]}\right)
\left[ \left(T^{[\ti\tj\tk]} T^{[\tl\tm\tn]} \right)-9\left(T^{[\ti\tj\tl]}T^{[\tk\tm\tn]}\right)\right]
\label{Riccitensor}
\eea
and the corresponding Ricci scalar, $R=R_{db}\eta^{db}$, which is then
\bea
R &=& - \frac{1}{3!} \OM^{f[\ti\tj\tk\tl]} \OM_{f[\ti\tj\tk\tl]} \nn \\
&&+\frac{1}{3(4!)}\eta\e^{\tg_1\ldots\tg_8}\fp_{ca}{}_{[\tg_1\tg_2]}
\left(T_{[\tg_3\tg_4\tg_5]} \s^{ca} T_{[\tg_6\tg_7\tg_8]}\right) \nn \\
&&+\frac{1}{3(4!)}\bar{\eta}\e_{\tg_1 \ldots \tg_8}\fm_{ca}{}^{[\tg_1\tg_2]}
\left(T^{[\tg_3\tg_4\tg_5]} \bs^{ca}T^{[\tg_6\tg_7\tg_8]} \right) \nn \\
 & & + \frac{1}{8(4!)^2} \d_{\th_1}^{\tg_1}{}_{\ldots}^{\ldots}{}_{\th_5}^{\tg_5}
\left( T_{[\tf\tg_1\tg_2]} T_{[\tg_3\tg_4\tg_5]}
\right) \left( T^{[\tf\th_1\th_2]} T^{[\th_3\th_4\th_5]} \right) \nn \\
&&+\frac{1}{16(4!)} \left(T_{[\ti\tj\tk]}T_{[\tl\tm\tn]}\right)
\left[\left(T^{[\ti\tj\tk]} T^{[\tl\tm\tn]} \right)-9\left(T^{[\ti\tj\tl]} T^{[\tk\tm\tn]} \right)\right].
\label{Ricciscalar}
\eea

The $SU(8)$ curvature components are the following:
\bea
R_\d^\td{}_\g^\tc{}^\tb{}_\ta &=& \frac{\eta}{4!}
\eps_{\d\g}\e^{\td\tc\tb\te\tf\ti\tj\tk}(T_{[\ta\te\tf]}T_{[\ti\tj\tk]})
-\frac{\eta}{(3!)^2}\d_\ta^{(\td} \e^{\tc)\tb\tf_1...\tf_6}T_{[\tf_1\tf_2\tf_3]\d}T_{[\tf_4\tf_5\tf_6]\g}\,,
\\[2mm]
R_\td^\dd{}_\tc^\dg{}^\tb{}_\ta &=&-\frac{\bar{\eta}}{4!}
\eps^{\dd\dg}\e_{\td\tc\tb\te\tf\ti\tj\tk}(T^{[\ta\te\tf]}T^{[\ti\tj\tk]})
+\frac{\bar{\eta}}{(3!)^2}\d^\tb_{(\td} \e_{\tc)\ta\tf_1...\tf_6}T^{[\tf_1\tf_2\tf_3]\dd}T^{[\tf_4\tf_5\tf_6]\dg}\,,
\\[2mm]
R_\d^\td{}_\tc^\dg{}^\tb{}_\ta &=& 2 \vi (\d_\tc^\td T_\d^\tb{}_\ta^\dg
-\d_\ta^\tb U_\d^\td{}_\tc^\dg - \d_\tc^\td U_\d^\tb{}_\ta^\dg)
+4\vi(\d_\ta^\td U_\d^\tb{}_\tc^\dg +  \d_\tc^\tb U_\d^\td{}_\ta^\dg)
-T^{[\td\tb\tf]\dg}T_{[\tc\ta\tf]\d}\,.
\eea
\bea
R_\d^\td{}_c{}^\tb{}_\ta &=& \frac{1}{18}\left(\d ^\td_\ta{}^\ti_\te{}^\tj_\tf{}^\tk_\tg
\OM_c{}^{[\tb\te\tf\tg]}
- 3 \d_\ta^\tb \OM_c{}^{[\td\ti\tj\tk]}\right) T_{[\ti\tj\tk]\d}\,, \\
R_\td^\dd{}_c{}^\tb{}_\ta &=&
-\frac{1}{18}\left(\d_\td^\tb{}
_\ti^\te{}_\tj^\tf{}_\tk^\tg \OM_{c[\ta\te\tf\tg]} - 3 \d_\ta^\tb \OM_{c[\td\ti\tj\tk]}\right)
T^{[\ti\tj\tk]\dd}\,.
\eea
\be
R_{dc}{}^\tb{}_\ta =  \frac{1}{3} \left( \OM_d{}^{[\tb\ti\tj\tk]} \OM_{c[\ta\ti\tj\tk]} -
\OM_c{}^{[\tb\ti\tj\tk]} \OM_{d[\ta\ti\tj\tk]}
\right) \,.
\ee

%===================================================================================
\subsection*{Spinorial derivatives}
%===================================================================================
The expression of the spinorial derivatives of the basic superfields are obtained in general using the
results of the Bianchi identities as well as the algebra of covariant derivatives and they are called also
constituency equations. Let us sum up them here.

Spinorial derivatives of the scalars:
\bea
\cd_\vf^\tf T^{[\td\tc]\au} &=& -\frac{\eta}{2 (3!)}
\e^{\tf\td\tc\ti\tj\tk\tl\tm} T_{[\tk\tl\tm]\vf} T_{[\ti\tj]}{}^\au
\\
\cd^\df_\tf T^{[\td\tc]\au} &=& -\frac{1}{2(3!)}
\d_\tf^\td{}_\tk^\tc{}_\tl^\ti{}_\tm^\tj T^{[\tk\tl\tm]\df}T_{[\ti\tj]}{}^\au
\\
\cd_\df^\tf T_{[\td\tc]}{}^\au &=& -\frac{1}{2(3!)}
\d^\tf_\td{}^\tk_\tc{}^\tl_\ti{}^\tm_\tj T_{[\tk\tl\tm]\vf}T^{[\ti\tj]^\au}
\\
\cd^\df_\tf T_{[\td\tc]}{}^\au &=& -\frac{\bar{\eta}}{2(3!)}
\e_{\tf\td\tc\ti\tj\tk\tl\tm} T^{[\tk\tl\tm]\df} T_{[\ti\tj]\au}
\eea
Notice, that the spinorial derivatives of the scalars $T$ are contained by definition in the spinorial
components of the one--form $\OM\ =\ (DT)S$. Therefore, the above expressions are equivalent to the
relations
\be
\cd_\vf^\tf T\ =\ \OM_\vf^\tf T\,,\qquad\cd^\df_\tf T\ =\ \OM^\df_\tf T\,,
\ee
with $\OM_\vf^\tf$ and $\OM^\df_\tf$ given by the equations \equ{om1} and \equ{om2}.

Spinorial derivatives of the gravigini superfields:
\bea
\cd_\vf^\tf T_{[\tc\tb\ta]\a} &=& -4 \d_\tc^\tf{}_\tb^\ti{}_\ta^\tj
(\s^{ba}\eps)_{\vf\a} \cf_{ba[\ti\tj]}^+ + \frac{\bar{\eta}}{4!} \eps_{\vf\a}
\e_{\tc\tb\ta\ti\tj\tk\tl\tm} \left( T^{[\tf\ti\tj]}T^{[\tk\tl\tm]}\right)\label{derspin1}
\\[2mm]
\cd^\df_\tf T_{[\tc\tb\ta]\a} &=& -4 \vi (\bs^f\eps)^\df{}_\a
\OM_{f[\tf\tc\tb\ta]}
\\[2mm]
\cd_\vf^\tf T^{[\tc\tb\ta]\da} &=& -4 \vi (\s^f\eps)_\vf{}^\da
\OM_f{}^{[\tf\tc\tb\ta]}
\\[2mm]
\cd^\df_\tf T^{[\tc\tb\ta]\da} &=& -4 \d^\tc_\tf{}^\tb_\ti{}^\ta_\tj
(\bs^{ba}\eps)^{\df\da} \cf_{ba}^-{}^{[\ti\tj]} + \frac{\eta}{4!}
\eps^{\df\da} \e^{\tc\tb\ta\ti\tj\tk\tl\tm} \left( T_{[\tf\ti\tj]}
T_{[\tk\tl\tm]} \right)\label{derspin2}
\eea

Spinorial derivatives of $\OM_a$, the field strength of the scalars:
\bea
\cd_\vf^\tf \OM_e^{[\td\tc\tb\ta]} &=& -\frac{\eta}{12}
\e^{\tf\td\tc\tb\ta\ti\tj\tk} \cd_e T_{[\ti\tj\tk]\vf}
+ \frac{\vi}{3} \d_\te^\td{}_\ti^\tc{}_\tj^\tb{}_\tk^\ta
\left(\s^f T^{[\ti\tj\tk]}\right)_\vf F_{fe}{}^{[\tf\te]}
\nn\\[2mm]
 & & - \frac{\eta}{12}\e^{\te\td\tc\tb\ta\ti\tj\tk} \left[
T_{[\ti\tj\tk]\vf} T_e^\tf{}_\te +2 \left(\s_{ef} T_{[\ti\tj\tk]} \right)_\vf U^f{}^\tf{}_\te \right]
\\[2mm]
\cd^\df_\tf \OM_e^{[\td\tc\tb\ta]} &=& - \frac{1}{12}
\d_\tf^\td{}_\ti^\tc{}_\tj^\tb{}_\tk^\ta \cd_e
T^{[\ti\tj\tk]\df} + \frac{\vi\eta}{3} \e^{\te\td\tc\tb\ta\ti\tj\tk}
\left(\bs^f T_{[\ti\tj\tk]}\right)^\df F_{fe[\tf\te]}
\nn\\[2mm]
 & & +\frac{1}{12} \d_\te^\td{}_\ti^\tc{}_\tj^\tb{}_\tk^\ta
\left[ T^{[\ti\tj\tk]\df} T_e^\te{}_\tf + 2 \left( \bs_{ef}
T^{[\ti\tj\tk]} \right)^\df U^f{}^\te{}_\tf \right]
\\[2mm]
\cd_\vf^\tf \OM_{e[\td\tc\tb\ta]} &=& - \frac{1}{12}
\d^\tf_\td{}^\ti_\tc{}^\tj_\tb{}^\tk_\ta \cd_e
T_{[\ti\tj\tk]\vf} + \frac{\vi \eta}{3} \e_{\te\td\tc\tb\ta\ti\tj\tk}
\left(\s^f T^{[\ti\tj\tk]}\right)_\vf F_{fe}{}^{[\tf\te]}
\nn \\[2mm]
 & & -\frac{1}{12} \d^\te_\td{}^\ti_\tc{}^\tj_\tb{}^\tk_\ta
\left[ T_{[\ti\tj\tk]\vf} T_e^\tf{}_\te + 2 \left( \s_{ef}
T_{[\ti\tj\tk]} \right)_\vf U^f{}^\tf{}_\te \right]
\\[2mm]
\cd^\df_\tf \OM_{e[\td\tc\tb\ta]} &=& -\frac{\bar{\eta}}{12}
\e_{\tf\td\tc\tb\ta\ti\tj\tk} \cd_e T^{[\ti\tj\tk]\df}
+ \frac{\vi}{3} \d^\te_\td{}^\ti_\tc{}^\tj_\tb{}^\tk_\ta
\left(\bs^f T_{[\ti\tj\tk]}\right)^\df F_{fe[\tf\te]}
\nn\\[2mm]
 & & + \frac{\bar{\eta}}{12} \e_{\te\td\tc\tb\ta\ti\tj\tk}\left[
T^{[\ti\tj\tk]\df} T_e^\te{}_\tf +2 \left(\bs_{ef} T^{[\ti\tj\tk]} \right)^\df U^f{}^\te{}_\tf \right]
\eea

Spinorial derivatives of the self--dual and anti--self--dual super--covariant field strength:
\bea
\cd_\vf^\tf \fm_{ba}{}^{[\tb\ta]} &=& \frac{\vi}{8} (\s^f\bs_{ba})_{\vf\da}
\left[\cd_f T^{[\tf\tb\ta]\da}
+2 T^{[\tg\tb\ta]\da}  U_f{}^\tf{}_\tg
+\oint^{\tf\tb\ta} T^{[\tg\tb\ta]\da} (T-U)_f{}^\tf{}_\tg\right]
\\[2mm]
\cd_\tf^\df \fm_{ba}{}^{[\tb\ta]} &=&
(\eps \bs_{ba})_{\db\da} \d^{[\tb}_\tf \S^{(\df\db\da)\ta]}
-\frac{\vi}{8}\left[\OM_f{}^{[\tb\ta\ti\tj]}\d^\tk_\tf
+\frac{1}{9}\d_\tf^\tb{}_\tg^\ta\OM_f{}^{[\tg\ti\tj\tk]}\right]
\left(\bs_{ba}\bs^f T_{[\ti\tj\tk]} \right)^\df
\nn \\[2mm]
&&+\frac{\bar{\eta}}{3!(4!)^2} \e_{\tf\te\ti\tj\tk\tl\tm\tn}
\left[\left(T^{[\ti\tj\tk]}\bs_{ba}T^{[\tl\tm\tn]}\right)T^{[\te\tb\ta]\df}
-2 \left(T^{[\ti\tj\tk]}\bs_{ba}T^{[\te\tb\ta]}
\right) T^{[\tl\tm\tn]\df} \right]
\nn \\[2mm]
&&+\frac{\bar{\eta}}{4(4!)^2}
\left[\frac{1}{9}\d^{\tb\ta}_{\tf\tn}\d^{\td\tc}_{\tk\tl}
+\d^{\tb\ta}_{\tk\tl}\d^{\td\tc}_{\tf\tn}\right]
\e_{\td\tc\tm\ti\tj\tf_1\tf_2\tf_3}
(T^{[\tn\tl\tm]}T^{[\tf_1\tf_2\tf_3]}) (\bs_{ba}T^{[\ti\tj\tk]})^\df
\eea
\bea
\cd^\tf_\vf \fp_{ba}{}_{[\tb\ta]} &=&
(\eps \s_{ba})^{\b\a} \d_{[\tb}^\tf \S_{(\vf\b\a)\ta]}
-\frac{\vi}{8}\left[\OM_f{}_{[\tb\ta\ti\tj]}\d_\tk^\tf
+\frac{1}{9}\d^\tf_\tb{}^\tg_\ta\OM_f{}_{[\tg\ti\tj\tk]}\right]
\left(\s_{ba}\s^f T^{[\ti\tj\tk]} \right)_\vf
\nn \\[2mm]
&&+\frac{{\eta}}{3!(4!)^2} \e^{\tf\te\ti\tj\tk\tl\tm\tn}
\left[\left(T_{[\ti\tj\tk]}\s_{ba}T_{[\tl\tm\tn]}\right)T_{[\te\tb\ta]\vf}
-2 \left(T_{[\ti\tj\tk]}\s_{ba}T_{[\te\tb\ta]}
\right) T_{[\tl\tm\tn]\vf} \right]
\nn \\[2mm]
&&+\frac{{\eta}}{4(4!)^2}
\left[\frac{1}{9}\d_{\tb\ta}^{\tf\tn}\d_{\td\tc}^{\tk\tl}
+\d_{\tb\ta}^{\tk\tl}\d_{\td\tc}^{\tf\tn}\right]
\e^{\td\tc\tm\ti\tj\tf_1\tf_2\tf_3}
(T_{[\tn\tl\tm]}T_{[\tf_1\tf_2\tf_3]}) (\s_{ba}T_{[\ti\tj\tk]})_\vf
\\[2mm]
\cd^\df_\tf \fp_{ba}{}_{[\tb\ta]} &=&
\frac{\vi}{8} (\bs^f\s_{ba})^{\a\df}
\left[\cd_f T_{[\tf\tb\ta]\a}
-2 T_{[\tg\tb\ta]\a}  U_f{}^\tg{}_\tf
-\oint_{\tf\tb\ta} T_{[\tg\tb\ta]\a} (T-U)_f{}^\tg{}_\tf\right]
\eea
\end{appendix}

\end{document}